\documentclass[a4paper,11pt]{article}

\usepackage[utf8]{inputenc}
\usepackage[T1]{fontenc}

\usepackage{jcapmod}

\usepackage{amsmath}
\usepackage{amssymb}
\usepackage{physics}
\usepackage{mathrsfs}
\usepackage{bbold}

\usepackage{hyperref}

\usepackage{graphicx}

\usepackage[a4paper,margin=2.5cm]{geometry}

\numberwithin{equation}{section}

\begin{document}

\pagenumbering{roman}

\begin{titlepage}

\baselineskip=15.5pt \thispagestyle{empty}

\begin{center}
    {\fontsize{20.74}{24}\selectfont \bfseries Quantum Circuit Complexity\\ of Primordial Perturbations}
\end{center}

\vspace{0.1cm}
\begin{center}
    {\fontsize{12}{18}\selectfont Jean-Luc Lehners$^{1}$ and Jerome Quintin$^{1}$}
\end{center}

\begin{center}
\vskip8pt
\textsl{$^1$ Max Planck Institute for Gravitational Physics (Albert Einstein Institute),\\
Am M\"uhlenberg 1, D-14476 Potsdam, Germany}
\end{center}

\vspace{1.2cm}
\hrule \vspace{0.3cm}
\noindent {\bf Abstract}\\[0.1cm] 
We study the quantum circuit complexity of cosmological perturbations in different models of the early universe. A natural measure for the complexity of cosmological perturbations is based on the symplectic group, allowing us to identify complexity with geodesics in the hyperbolic plane. We investigate the complexity of both the mode functions and the physical perturbations, arguing that the latter often provides a more insightful description of the physics involved. In all models the total complexity reached is rather large. Inflationary perturbations may be represented by a comparatively simple quantum circuit, while the perturbations during a matter-dominated contracting phase present the most rapid growth in complexity. Ekpyrotic perturbations reside in the middle and are distinguished by the smallest growth of complexity before horizon exit. Our analysis serves to highlight how different cosmological models achieve the same end result for the perturbations via different routes and how all models show a pronounced sensitivity to initial conditions. 
\vskip10pt
\hrule
\vskip10pt

\end{titlepage}

\pagenumbering{arabic}
\thispagestyle{empty}
\setcounter{page}{2}
\tableofcontents

\section{Introduction}

The oldest optical record that we have of the universe is the cosmic microwave background radiation (CMB), which shows us the state of the universe about 380,000 years after the start of the hot big bang expansion. The CMB contains temperature fluctuations that, using known plasma physics, can be extrapolated back in time to primordial density fluctuations with an almost scale-invariant spectrum over the range of wavelengths that can be observed. In these calculations, it is an excellent approximation to treat the primordial perturbations as classical density perturbations. 

A striking idea is that the ultimate origin of the primordial perturbations lies in quantum fluctuations that were amplified and rendered effectively classical in the early universe. The best known scenario of this type is inflation \cite{Guth:1980zm,Linde:1981mu,Mukhanov:1981xt,Albrecht:1982wi}, which can simultaneously amplify quantum perturbations, render them classical and explain their seemingly acausal correlations by tracing them to earlier causal processes. It came as a surprise to many that there exist alternative scenarios, in particular ekpyrotic cosmology \cite{Khoury:2001wf,Lehners:2008vx} and a contracting matter phase \cite{Wands:1998yp}, that can achieve the same goal with entirely different physics. In all cases, however, the underlying idea is that quantum fluctuations are turned into effectively classical density perturbations. Thus in all these models the universe acts as a quantum computer, processing an initial quantum state (usually taken to be the vacuum state) into a state that can explain what we observe in the CMB. In this paper we want to explore this alternative description in terms of quantum computation, in particular by calculating the complexity of the involved computation. As the name suggests, the complexity of a quantum computation may be thought of as the difficulty in building a quantum computer performing the same task. More precisely, the complexity is taken to be the minimum number of quantum gates (from a specified set) that the task requires.  In other words, we are asking how complicated a quantum computer would have to be in order to simulate the perturbations in various early universe models.\footnote{Describing what the actual quantum computer or circuit would have to be to perform such a quantum simulation is not the goal of this paper though. For a first attempt at developing a quantum algorithm for inflation, see \cite{Li:2020kbv}.}

\begin{figure}[t]
\centering
\includegraphics[width=0.35\textwidth]{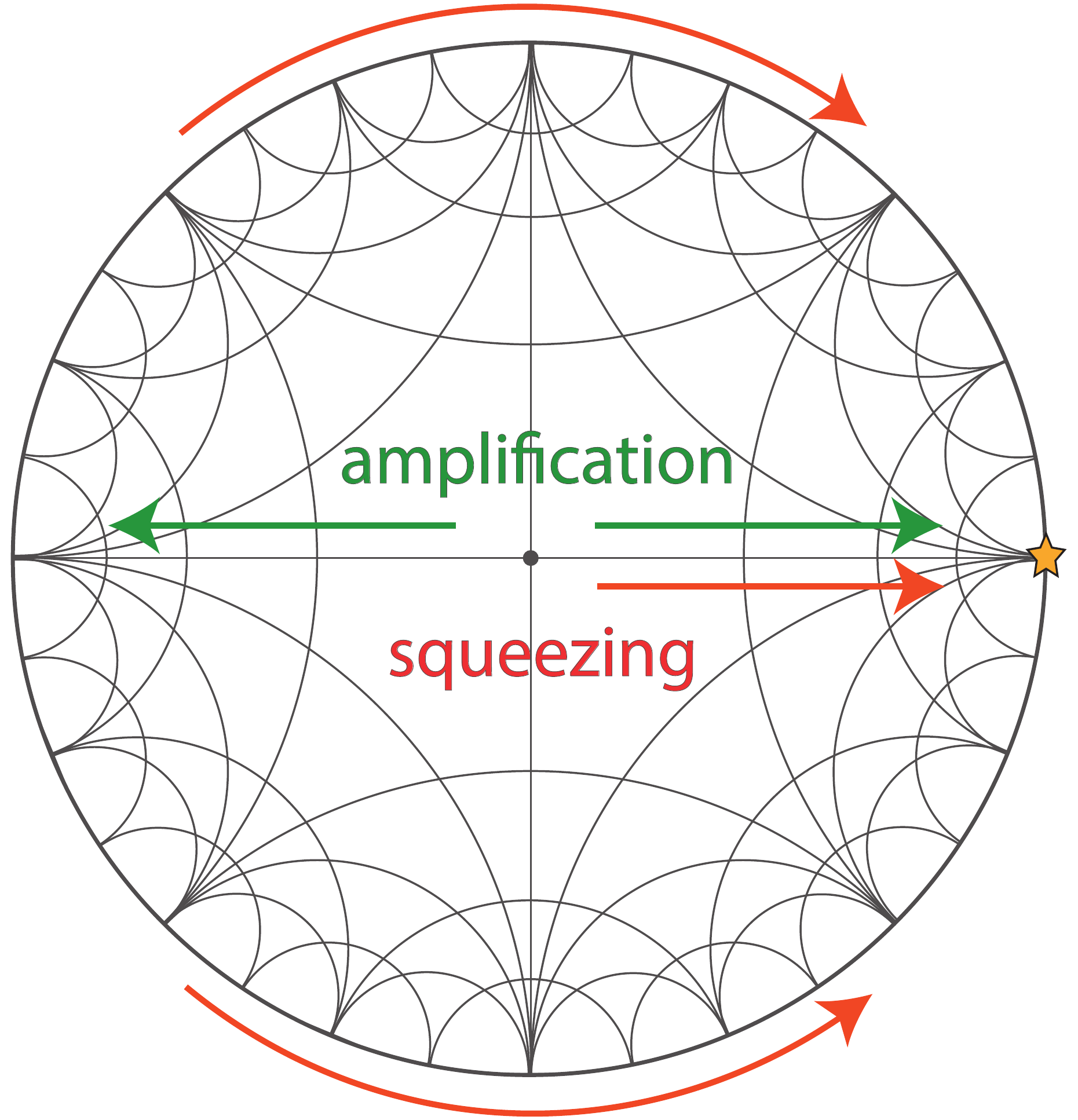}
\caption{We may usefully display the evolution of the quantum state of cosmological perturbations on a Poincar\'{e} disk, which forms a representation of hyperbolic geometry. Starting from the Bunch-Davies vacuum state in the centre of the disk, the arrows indicate the directions of evolution in which the perturbations are amplified or squeezed. A final state that is both amplified and squeezed, as required to match the observations of the CMB, corresponds to a state very close to the boundary directly right from the centre, as indicated by the star.} \label{fig:PoincareDiskGen}
\end{figure}

\begin{figure}[t]
\centering
\includegraphics[width=\textwidth]{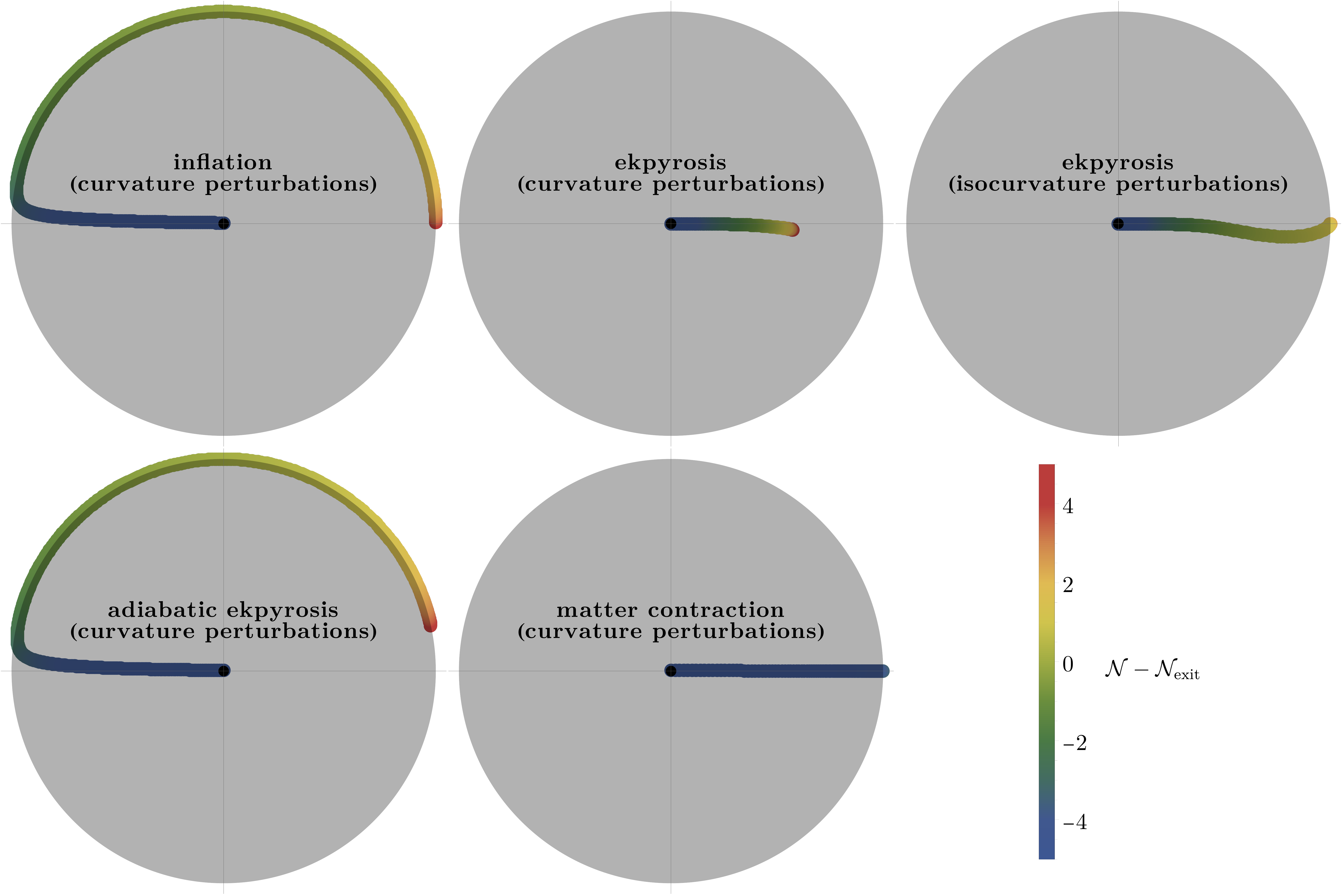}
\caption{The evolution of cosmological perturbations represented on the Poincar\'{e} disk (cf.~Fig.~\ref{fig:PoincareDiskGen}). The colour code depicts the time evolution as a function of $e$-folding number $\mathcal{N}$ for a given mode from its sub-horizon, ultraviolet regime (in dark blue), through horizon exit at $\mathcal{N}=\mathcal{N}_\mathrm{exit}$ (in chartreuse colour), all the way to its super-horizon, infrared regime (in red). The differences between inflation (where amplification and squeezing occur in succession) with, for instance, isocurvature perturbations during ekpyrosis (where amplification and squeezing occur simultaneously) become manifest. Not all perturbations achieve a highly amplified and squeezed end state: in particular, curvature perturbations during ekpyrosis do not, as the evolution does not get close to the boundary. Meanwhile, a contracting matter phase reaches the boundary extremely fast, with much of the distance covered already sub-horizon.} \label{fig:PoincareDisk}
\end{figure}

As we will see, for the states relevant to early universe cosmology, we will quite naturally be led to depict their evolution in hyperbolic geometry; see Fig.~\ref{fig:PoincareDiskGen}. The cosmological phases mentioned above turn quantum fluctuations into effectively classical fluctuations by combining two effects: they amplify the perturbations and also turn the quantum state into a highly squeezed state where the uncertainty in momentum is vastly smaller than that in amplitude. In this manner the quantum states become equivalent to a statistical mixture of classical perturbations \cite{Grishchuk:1990bj,Brandenberger:1990bx,Albrecht:1992kf,Grishchuk:1993ds,Polarski:1995jg,Prokopec:2006fc,Kiefer:2007zza,Martin:2012ua,Battarra:2013cha}. On the Poincar\'{e} disk in Fig.~\ref{fig:PoincareDiskGen} we have indicated by arrows the directions of evolution corresponding to amplification and squeezing, starting from the vacuum state in the centre. Even though all viable early universe models must end up with the same final state, they employ rather different routes to get there. Figure \ref{fig:PoincareDisk} illustrates this by showing the evolution in the different early universe models that we investigate in this paper (the actual numerical calculations performed to produce this figure will be explained in due course). As is immediately apparent from the figure, in some models amplification and squeezing occur separately (in particular in inflation), while in other models they occur simultaneously (e.g.~for isocurvature perturbations during ekpyrosis). Thus it may already be guessed that the complexities of the corresponding computations will also turn out to be different, and this is indeed what we find. 

In fact, we find that the complexity depends quite strongly on the cosmological model. In all cases the final complexity is very high, due to the vast range of scales that need to be processed. We actually find that it is useful to characterise the evolution of complexity in terms of the number of $e$-folds, rather than in terms of physical time. Broadly speaking, inflation turns out to be a simpler quantum computer than contracting cosmologies, with a contracting matter phase being the most complex. Ekpyrotic perturbations behave in an intermediate regime, though they have the feature that complexity grows the slowest while perturbations are still on sub-horizon scales. An interesting aspect is that the complexity is sensitive to the total duration, and thus also to the beginning, of the cosmological phases in question. This offers the hope that complexity may help in further elucidating the conditions required at the beginning of the early universe models considered, in order to see how they may eventually form a part of a complete cosmology.

We will start the paper with two review sections, one on circuit complexity (section \ref{sec:reviewQCC}) and one on the quantisation of cosmological perturbations (section \ref{sec:quantisation}). Readers may skip them if they feel comfortable with the subjects, though we review them with a particular application in mind, and hence the approach may be interesting even for the expert reader. We will then apply these methods to investigate circuit complexity for different cosmological models. Inflation is considered in section \ref{sec:inflation}, while we divide the analysis for ekpyrosis up into single-field models (section \ref{sec:ekpyrosis}) and two-field models (section \ref{sec:twofields}), given that the involved physics is significantly different. We then compare these results with matter contraction and adiabatic ekpyrosis in section \ref{sec:matter}, where we will see that these models provide the extremes on the spectrum of complexity. Our results are discussed in section \ref{sec:discussion}, and we include a brief comparison with earlier proposals in appendix \ref{app:comparison}. We use natural units with $\hbar=1$ and $8\pi G_\mathrm{N}=1$ throughout.

\section{Brief review of quantum circuit complexity}\label{sec:reviewQCC}

In quantum computation, complexity describes how difficult it is to build a circuit that transforms a given reference state $|\Psi_\mathrm{R}\rangle$ into a target state $|\Psi_\mathrm{T}\rangle$. A conceptually straightforward measure for the difficulty of performing a computational task is simply to look at how many quantum gates one needs in order to perform the required transformation. Complexity thus provides a quantitative way of evaluating how much a wave function has changed. The fact that we will only be interested in Gaussian perturbations greatly simplifies the analysis (a useful exposition, which we will partly follow here, can be found in \cite{Jefferson:2017sdb}). Let us assume in the present section that we are working with a one-dimensional harmonic oscillator in position space $x$. We will further assume that the reference and target states are given by the wave functions
\begin{equation}
 |\Psi_\mathrm{R}\rangle = \left(\frac{\omega}{\pi}\right)^{1/4} \, e^{-\frac{1}{2}\omega x^2}\,, \qquad |\Psi_\mathrm{T}\rangle =\left(\frac{\Omega}{\pi}\right)^{1/4}\, e^{-\frac{1}{2}\Omega x^2}\,, \label{eq:referencetarget}
\end{equation}
with frequencies $\omega$ and $\Omega$, which at first we take to be positive real numbers.
The evolution between the reference and target states will be unitary,
\begin{equation}
 |\Psi_\mathrm{T}\rangle = U |\Psi_\mathrm{R}\rangle\,. \label{evol}
\end{equation}
The question is then how many gates one needs in order to implement the unitary operator~$U$ or, rather, what the minimum number of such gates might be. We can proceed by first discretising the evolution by considering small steps of size $\epsilon$. Our gates will be elementary unitary operators. Useful examples are given by the following operators, displayed here along with the effect that they have on a wave function $|\Psi(x)\rangle$:
\begin{subequations}
\begin{align}
 & H \equiv e^{i\epsilon}\,, & H|\Psi(x)\rangle = e^{i\epsilon} |\Psi(x)\rangle\,; \\
 & J \equiv e^{i\epsilon p_x} = e^{\epsilon \partial_x}\,, & J |\Psi(x)\rangle = |\Psi(x+\epsilon)\rangle\,; \\
 & Q \equiv e^{\frac{\epsilon}{2}} e^{i \epsilon x p_x} = e^{\frac{\epsilon}{2}} e^{\epsilon x \partial_x}\,, & Q |\Psi(x)\rangle = e^{\frac{\epsilon}{2}} |\Psi(e^\epsilon x)\rangle\,.
\end{align}
\end{subequations}
Here $p_x$ denotes the momentum operator conjugate to $x,$ that is to say $p_x = -i \partial_x.$ In the examples above, $H$ effects a phase change in the wave function, while $J$ shifts the position. Most useful to us is $Q,$ which leads to a scaling in the position (and also includes a normalisation factor). These are just a few examples --- one could in principle consider many more operators. 

A general circuit is built from a number of such gates performed in succession, i.e.,
\begin{equation}
 U = \cdots H^{\alpha_i} J^{\alpha_j} Q^{\alpha_k} H^{\alpha_\ell} J^{\alpha_m} Q^{\alpha_n}\cdots\,,
\end{equation}
where the $\alpha$s are natural numbers. General circuits would contain operations that are later undone again by other gates. We are interested in the minimal number of gates required, however. For our specific example in Eq.~\eqref{eq:referencetarget} it is clear that we will only need the scaling ($Q$) gate, so that for us a useful circuit will be of the form
\begin{equation}
 U = Q^\alpha \quad \longrightarrow \quad Q^\alpha e^{-\frac{1}{2}\omega x^2} = e^{\frac{\alpha \epsilon}{2}} e^{-\frac{1}{2}\omega e^{2\alpha \epsilon} x^2}\,.
\end{equation}
Thus we can obtain the required transformation from $|\Psi_\mathrm{R}\rangle$ to $|\Psi_\mathrm{T}\rangle$ as long as $2 \alpha \epsilon = \ln(\Omega/\omega)$. In general one should now assign a measure to the `depth' of the circuit. A popular method is to assign a metric to the space of unitary operations, after taking the continuum limit $\epsilon \to 0.$ This is useful, as the shortest circuit will then correspond to a geodesic in this space, and this geometric method allows one to use the power of differential geometry (see \cite{Nielsen2005}).
We will turn to such an example shortly.
Our present example is, however, so simple that it is sufficient to directly equate the complexity $\mathcal{C}$ to the number of gates, where we have to re-scale the measure by $\epsilon$ in order to obtain a well-defined limit as $\epsilon \to 0$:
\begin{equation}
 \mathcal{C} = \epsilon \alpha = \frac{1}{2}\ln\left(\frac{\Omega}{\omega}\right)\,. \label{complexityone}
\end{equation}
Note that the complexity evolves logarithmically; hence even small numerical changes in the complexity correspond to significant evolution of the wave function. As we will see, the early universe provides a laboratory that is surprisingly efficient at producing complex quantum circuits.

In cosmology, the frequencies $\omega,\,\Omega$ are in general complex valued since we are dealing with generic Gaussian states. We therefore have to find an appropriate generalisation of Eq.~\eqref{complexityone}. Perhaps the most straightforward generalisation is simply to let $\alpha$ become a complex number, too, which one should then think of as encoding the effect of two quantum gates, one for the real part and one for the imaginary part of $\alpha$. This has been formalised in \cite{Ali:2018fcz} (and used in the cosmological context; see, e.g., \cite{Bhattacharyya:2020rpy,Bhattacharyya:2020kgu}), with the complexity now defined via
\begin{equation}
 \mathcal{C}^{\mathrm{(a.c.)}} = \epsilon |\alpha| = \frac{1}{2}\left|\ln\left(\frac{\Omega}{\omega}\right)\right| = \frac{1}{2}\sqrt{\left(\ln\left|\frac{\Omega}{\omega}\right|\right)^2+\left(\arctan\left[\frac{\Im(\Omega/\omega)}{\Re(\Omega/\omega)}\right]\right)^2}\,.\label{eq:defcomplexity}
\end{equation}
We will refer to this definition as the analytically continued (a.c.) complexity.

There exists another generalisation, however, which leads to an appealing geometric interpretation, both of the states and of their complexity. This generalisation was developed in \cite{Camargo:2018eof,Chapman:2018hou} and takes as its starting point the covariance matrices associated with Gaussian states. Let us first rewrite the target state as
\begin{equation}
 |\Psi_\mathrm{T}\rangle  = \left( \frac{a}{\pi}\right)^{1/4}\, e^{-\frac{1}{2}(a+ib)x^2}\,,
\end{equation}
with $a,\,b$ being real valued. Moreover, we group the position and momentum into a combined coordinate $\xi^m=(x,p)$ with $m=1,2$. Since we are working with a Gaussian state, all information is contained in the quadratic combinations 
\begin{equation}
 2\langle \Psi_\mathrm{T} | \xi^m \xi^n | \Psi_\mathrm{T} \rangle = G_\mathrm{T}^{mn} + i \Omega^{mn}\,,
\end{equation}
where the antisymmetric matrix 
\begin{equation}
 \Omega = \begin{pmatrix}
   0 & 1\\ 
  -1 & 0
 \end{pmatrix}
\end{equation}
encodes the canonical commutation relations, while the symmetric covariance matrix $G_\mathrm{T}$ has entries given by the expectation values
\begin{equation}
 G_\mathrm{T} = \begin{pmatrix}
  2\langle x^2 \rangle& \langle xp+px \rangle\\ 
  \langle xp+px \rangle & 2\langle p^2 \rangle
 \end{pmatrix}
 =
 \begin{pmatrix}
  \frac{1}{a}& -\frac{b}{a}\\ 
  -\frac{b}{a} & \frac{a^2 + b^2}{a}
 \end{pmatrix}\,. \label{covariance}
\end{equation}
The evolution equation \eqref{evol} now becomes 
\begin{equation}
 G_\mathrm{T} = U G_\mathrm{R} U^{T} \,. \label{evol2}
\end{equation}
The transformations $\tilde\xi^m = \tilde{M}^m{}_n \xi^n$ that preserve the canonical commutation relations are elements of the symplectic group, $\tilde{M} \in \mathrm{Sp}(2,\mathbb{R})$. This suggests the use of gates that belong to the algebra $\mathfrak{sp}(2)$. In fact, it is precisely quadratic combinations of $x$ and $p$ that form the corresponding generators,
\begin{subequations}
\begin{align}
 &V \equiv \frac{i}{\sqrt{2}}x^2\,, \qquad W \equiv \frac{i}{2}(xp+px)\,, \qquad Z \equiv \frac{i}{\sqrt{2}}p^2\,, \\
 &[V,W]=-2V\,,\qquad [V,Z]=-2W\,,\qquad [W,Z]=-2Z\,.
\end{align}
\end{subequations}
As noticed in \cite{Camargo:2018eof}, it is sufficient to consider the sub-algebra formed by $V$ and $W$ --- moreover, this restriction will lead to a useful connection with hyperbolic geometry, as we will now review.

A matrix representation of $V,\,W$, along with the associated group elements/gates is provided by
\begin{subequations}
\begin{align}
 &V_{2\times 2}=\begin{pmatrix}
  0& 0\\ 
  \sqrt{2} & 0
 \end{pmatrix}\,, \qquad W_{2\times 2}=\begin{pmatrix}
  -1& 0\\
  0 & 1
 \end{pmatrix}\,,\\
 &Q_V = e^{\epsilon V_{2\times 2}}\,, \qquad\qquad\ \, Q_W = e^{\epsilon W_{2\times2}}\,.
\end{align}
\end{subequations}
These gates are sufficient to take a reference matrix $G_\mathrm{R}=\mathrm{diag}(1/\omega,\omega)$ to the target \eqref{covariance}. We may slightly simplify the task by using a squeezing operation $S = \mathrm{diag}(\sqrt{\omega},1/\sqrt{\omega})$ on the reference and target states, with the effect that
\begin{equation}
 S G_\mathrm{R} S^T = \mathbb{1}\,, \qquad \tilde{G}_\mathrm{T} \equiv S G_\mathrm{T} S^T =  
 \begin{pmatrix}
  \frac{\omega}{a}& -\frac{b}{a}\\ 
  -\frac{b}{a} & \frac{a^2 + b^2}{\omega a}
 \end{pmatrix}\,.
\end{equation}
Thus the reference has become the identity.

A general circuit will again consist of a sequence of gates. It is here that we will make the transition from a discrete to a continuous description, which will allow us to make a connection with differential geometry. In the continuous description a circuit is represented by a path-ordered exponential
\begin{equation}
 U(s) = \overleftarrow{P}\exp\left(\int_0^s \dd\tilde{s}\,Y^I(\tilde{s}) M_I\right)\,, \label{continuouscircuit}
\end{equation}
with $M_I = (V_{2\times 2},W_{2\times 2})$ and $Y^I(s)$ describing switches that turn the respective gates on or off. The full circuit then runs from the identity at $s=0$ to the target $\tilde{G}_\mathrm{T}$ at $s=1.$ 

It is straightforward to verify that a general element of such a circuit can be parameterised by
\begin{equation}
 U = \begin{pmatrix}
  \sqrt{z}& 0\\ 
  \frac{y}{\sqrt{2z}} & \frac{1}{\sqrt{z}}
 \end{pmatrix}\,.
\end{equation}
Note that the transformation law \eqref{evol2} involves the transpose, thus also generating the required first row/second column entry.  The connection to geometry is found by inverting \eqref{continuouscircuit} according to 
\begin{equation}
 \dd Y^I = \frac{1}{2}\mathrm{Tr}(\dd U\,U^{-1} {M_I}^T)\,,
\end{equation}
where the factor $1/2$ comes from the normalisation of the gates $M_I$. From this we can immediately obtain a circuit geometry specified by the line element
\begin{equation}
 \dd s^2 = g_{IJ} \dd Y^I \dd Y^J = \frac{\dd z^2 + \frac{1}{2} \dd y^2}{4 z^2}\,.
\end{equation}
Here, following \cite{Camargo:2018eof}, we chose $g_{IJ}=\mathrm{diag}(1,1/2)$, but other choices are equally simple to implement. The line element above may be recognised as the metric on the hyperbolic upper half plane $\mathbb{H}^2$. Optimal circuits then correspond to geodesics, which on the hyperbolic plane are given by arcs of circles that are perpendicular to the boundary $z=0.$ Our reference and target states correspond to the coordinate locations
\begin{equation}
 (y_0,z_0) = (0,1)\,, \qquad (y_1,z_1) = \left(-\frac{b}{\sqrt{2}a},\frac{\omega}{a}\right)\,.
\end{equation}
Then the complexity may be defined as the hyperbolic distance between these points,
\begin{subequations}\label{eq:defcomplexity2}
\begin{align}
 \mathcal{C} &= \frac{1}{2} \mathrm{argcosh}(X)=\frac{1}{2}\ln\left(X+\sqrt{X^2-1}\right)\,, \label{eq:defcomplexityCh2} \\
 X &= \frac{z_0^2 + z_1^2 + \frac{1}{2}(y_1-y_0)^2}{2z_0z_1} = \frac{1}{2}\left(\frac{a}{\omega} + \frac{\omega}{a} + \frac{b^2}{\omega a}\right) = \frac{1}{2}\, \mathrm{Tr} \, \tilde{G}_\mathrm{T}\,.
\end{align}
\end{subequations}
We will refer to this definition simply as the ``complexity'', though for the sake of comparison with the analytically continued complexity \eqref{eq:defcomplexity} (such as in appendix \ref{app:comparison} for instance) we shall sometimes specifically refer to it as the ``hyperbolic complexity''.

Finally, we mention that it can be useful to map the hyperbolic plane to a finite representation, in particular to the Poincar\'{e} disk. This mapping is most easily expressed in terms of the complex coordinate $Z \equiv y+ i z$. Then a point $Z$ on the hyperbolic half-plane is mapped to the point $\frac{Z-i}{Z+i}$ on the disk. In particular, our reference state $(0,1)$ corresponds to $Z=i$ and gets mapped to the origin of the disk. Target states that are far from the reference state, and thus obtain a large hyperbolic complexity, can then be found very close to the edge of the Poincar\'{e} disk. This is the representation that is used in Figs.~\ref{fig:PoincareDiskGen} and \ref{fig:PoincareDisk}. It thus becomes clear how amplification and squeezing of the perturbations ($\omega/a$ and $b/a$ growing, respectively) moves one around the Poincar\'e disk, getting closer to the edge and the point $(1,0)$ on the disk as $|Z|$ grows.

\section{Quantisation of cosmological perturbations and complexity thereof} \label{sec:quantisation}

The easiest way to model both inflation and ekpyrosis is by considering the dynamics of scalar fields coupled to gravity and evolving in an appropriate potential. We will start our analysis with models involving only a single scalar field $\sigma$ with an exponential potential $V(\sigma).$ The action is given by
\begin{equation}
 S = \int \dd^4x \, \sqrt{-g} \left( \frac{R}{2} - \frac{1}{2} (\partial\sigma)^2 - V_0 e^{-\sqrt{2\epsilon}\sigma} \right)\,.\label{eq:Ssinglefield}
\end{equation}
Note that $\epsilon$ is always taken to be positive, but $V_0$ can be positive or negative. In a flat Robertson-Walker spacetime, the equations of motion reduce to
\begin{subequations}
\begin{eqnarray}
3H^2 &=& \frac{1}{2}\dot\sigma^2 + V \, , \label{FRW1} \\
\dot{H} &=& - \frac{1}{2} \dot\sigma^2 \, ,\label{FRW2}\\
\ddot{\sigma} + 3 H \dot{\sigma} + \partial_\sigma V &=& 0 \, .\label{FRW3}
\end{eqnarray}
\end{subequations}
When $\epsilon$ is constant [i.e., when the scalar field equation of state (EoS) is constant], there exists an exact scaling solution to the  equations of motion, given by
\begin{equation} \label{ScalingSolution}
 a = a_0 |t|^{1/\epsilon}\,, \quad H = \frac{1}{\epsilon t}\,, \quad \sigma = \frac{1}{\sqrt{2\epsilon}}\ln \left( \frac{V_0\, \epsilon^2}{3-\epsilon} \, t^2\right)\,, \quad V = \frac{3-\epsilon}{\epsilon^2 t^2}\,.
\end{equation}
For inflation, one would take $V_0>0$ and assume the universe to be expanding (i.e., $t>0$), with EoS $\epsilon < 1$. This then corresponds to an accelerated expansion of the scale factor and slow evolution of the scalar field. For ekpyrosis, one would take $V_0<0$ and assume the universe to be contracting (i.e., $t < 0$) with $\epsilon > 3$. This would correspond to slow contraction and fast evolution of the scalar field. Note that the EoS (slow-roll/fast-roll) parameter $\epsilon$  can be re-expressed in various ways,
\begin{equation}
 \epsilon = - \frac{\dot{H}}{H^2} = \frac{1}{2}\frac{\dot\sigma^2}{H^2} = \frac{1}{2}\frac{(\partial_\sigma V)^2}{V^2} \,.
\end{equation}
Below, we will work in conformal time $\tau$, defined by $\dd\tau\equiv\dd t/a$, in terms of which the scaling solution becomes
\begin{equation}
 (\epsilon - 1) \tau = \epsilon \, t^{-(1-\epsilon)/\epsilon}\,, \quad a(\tau)=\bar{a}_0 (-\tau)^{\frac{1}{\epsilon - 1}}\,, \quad \mathcal{H} \equiv \frac{a'}{a} = \frac{1}{(\epsilon - 1)\tau}\,, \label{eq:backscal}
\end{equation}
where a prime denotes a derivative with respect to conformal time. Conformal time naturally runs over negative values for both inflation and ekpyrosis.
We will also often use the $e$-folding number to characterise the time dependence later.
We define the number of $e$-folds of evolution via $\dd\mathcal{N} \equiv \dd\ln(a|H|)=\dd\ln|\mathcal{H}|$ since it is the appropriate definition with regard to the flatness problem.
For inflation $H$ is roughly constant so that instead one often uses $\dd N = \dd\ln a$, but this is not useful for ekpyrosis where $a$ is approximately constant.
Thus, we will use $\dd\mathcal{N}=-\dd\ln(-\tau)=(1-\epsilon)\dd\ln a$, which is valid generally.

We will be primarily interested in perturbations of these models. For this, we will work in comoving gauge, in which the scalar field fluctuation vanishes, $\delta\sigma=0$, and the scalar degree of freedom is represented by the comoving curvature perturbation $\mathcal R$. The metric on spatial hypersurfaces is then given by
\begin{equation}
 h_{ij}=a^2(1+2\mathcal R)\delta_{ij}\,.
\end{equation}
We will only consider scalar perturbations --- an analogous calculation could be performed for gravitational waves. At quadratic order, the action for the comoving curvature perturbation is remarkably simple and given by
\begin{equation}
 S^{(2)}=\int \mathrm{d}^3x \mathrm{d}t \, \epsilon \left(a^3 \dot{\mathcal{R}}^2-a(\partial_i \mathcal R)^2\right)\,. \label{ActionComovingCurvature}
\end{equation}
The corresponding equation of motion is
\begin{equation} \label{Reom}
 \ddot{\mathcal{R}} + \left( 3H + \frac{\dot\epsilon}{\epsilon}\right) \dot{\mathcal{R}} - \frac{1}{a^2}\partial_i\partial^i \mathcal{R} = 0\,.
\end{equation}

In order to quantise the perturbations, it is useful to define the Mukhanov-Sasaki variable
\begin{equation} \label{MSvariable}
 v\equiv z \mathcal R\,,\qquad z^2\equiv 2a^2 \epsilon=a^2\frac{\dot \phi^2 }{H^2}\,.
\end{equation}
Switching to conformal time, the action \eqref{ActionComovingCurvature} now becomes canonically normalised:
\begin{equation}
 S^{(2)}= \frac{1}{2} \int \mathrm{d}^3x \mathrm{d}\tau\,\left[(v')^2 -(\partial_i v)^2-2\frac{z'}{z}vv'+\left(\frac{z'}{z}\right)^2v^2\right]\,.
\end{equation}
The action is quadratic and will thus lead to a linear equation of motion. This implies that it will be useful to expand the perturbations into Fourier modes:
\begin{equation}
 v(\tau, \mathbf{x})=\int \frac{\mathrm{d}^3k}{(2\pi)^{3}} \, v_{\mathbf k}(\tau) e^{i \mathbf{k}\cdot\mathbf{x}}\,.
\end{equation}
The equation of motion for each Fourier mode of wavenumber $\mathbf{k}$ is then
\begin{equation}
 v''_{\mathbf{k}}+\left(k^2-\frac{z''}{z}\right)v_{\mathbf{k}}=0\,. \label{FourierEom}
\end{equation}
The linearity of the equation implies that each mode evolves independently, and there is no mode mixing.

We can quantise the system in the Heisenberg picture by promoting the mode functions to operators and writing these new operators as a linear combination of annihilation and creation operators,
\begin{equation}
 \hat{v}_{\mathbf{k}} = f_k(\tau) \hat{a}_{\mathbf{k}} + f^*_{k}(\tau) \hat{a}^\dagger_{-\mathbf{k}}\,.
\end{equation}
Here the $f_k(\tau)$ are time-dependent (complex) solutions of the equations of motion \eqref{FourierEom}, which because of the spatial isotropy of the background depend only on the modulus $k\equiv|\mathbf{k}|$. Note that the definition above implies the relation $\hat{v}_{-\mathbf{k}} = \hat{v}_{\mathbf{k}}^\dagger$, which ensures that the comoving curvature perturbation is real valued, as it should.
We then require the annihilation/creation operators to satisfy the canonical quantisation condition,
\begin{equation}
 [\hat{a}_{\mathbf{k}},\hat{a}^{\dagger}_{-\mathbf{p}}]=(2\pi)^3\delta^{(3)}(\mathbf{k} + \mathbf{p})\,.
\end{equation}
This condition implies that the field operator $\hat{v}$ and its conjugate momentum $\hat\pi \equiv \hat{v}^\prime - (z'/z)\hat{v}$ satisfy the canonical equal-time commutation relation,
\begin{equation}
 [\hat{v}(\tau,\mathbf{x}),\hat{\pi}(\tau,\mathbf{y})]=i\delta^{(3)}(\mathbf{x}-\mathbf{y})\,,
\end{equation}
as well as the trivial commutators,
\begin{equation}
 [\hat{v}(\tau,\mathbf{x}),\hat{v}(\tau,\mathbf{y})]=[\hat{\pi}(\tau,\mathbf{x}),\hat{\pi}(\tau,\mathbf{y})]=0\,.
\end{equation}
The Wronskian is a constant of motion,
\begin{equation} \label{Wronski}
 f_kf^{*\prime}_k-f^*_kf^\prime_k=i\,,
\end{equation}
where we have fixed the right-hand side in such a way as to ensure the canonical normalisation of the mode functions. 

The above quantisation procedure in the Heisenberg picture is standard in early universe cosmology. But in order to investigate the analogy with circuit complexity, it is more useful to work directly with the wave function, i.e., in the Schr\"{o}dinger picture. The wave function will be of Gaussian form since at our level of approximation the perturbations are governed by a quadratic action. It will, in fact, be a product of Gaussians for each wavenumber $k$, and thus we may focus on a single wavenumber and write
\begin{equation}
 |\Psi(v)\rangle \propto \exp\left(-\frac{1}{2}A_{\sigma\sigma}v^2\right) = \exp\left(-\frac{1}{2}A_{\mathcal{R}\mathcal{R}}\mathcal{R}^2\right)\,, \label{eq:wave function}
\end{equation}
where the proportionality constant is determined upon normalisation.
We are interested in the vacuum state $\hat{a}|\Psi\rangle = 0.$ If we rewrite the annihilation operator as $i \hat{a}=(f^{*\prime}-\frac{z'}{z}f^*)\hat{v} - f^* \hat{\pi},$ then with $\hat\pi \to -i\partial_v$ we can deduce an expression for the correlator $A_{\sigma\sigma}$ in terms of the mode functions as
\begin{equation}
 A_{\sigma\sigma} = - i \frac{f_{k}^{*\prime}}{f_{k}^*} + i \frac{z'}{z} \,, \qquad A_{\mathcal{R}\mathcal{R}} = z^2 A_{\sigma\sigma}\,.\label{eq:Corrsigma}
\end{equation}
We also added the correlator of the comoving curvature perturbation, whose expression follows immediately upon using $v=z\mathcal{R}$. The equation of motion for the correlator follows from the  Schr\"{o}dinger equation $i|\Psi\rangle' = \hat{H}|\Psi\rangle$, where $\hat{H}$ is the Hamiltonian operator, or equivalently from the Heisenberg equation of motion,
\begin{equation}
 i A_{\sigma\sigma}' = \left(A_{\sigma\sigma}-i\frac{z'}{z}\right)^2 - k^2 + \frac{z^{\prime 2}}{z^2}\,.
\end{equation}

We can now solve for the mode functions defined above. On a scaling solution, the mode equation \eqref{FourierEom} becomes 
\begin{equation}
 f_k'' + \left( k^2 - \frac{2-\epsilon}{(1-\epsilon)^2\tau^2} \right) f_k = 0\,.
\end{equation}
It can be solved exactly by rewriting it in the form
\begin{equation}
 f_k'' + \left(k^2 - \frac{\alpha_\sigma^2 - 1/4}{\tau^2} \right) f_k = 0\,, \qquad \textrm{where} \quad \alpha_\sigma \equiv \frac{1}{2}\left|\frac{3-\epsilon}{1-\epsilon}\right|\,.
\end{equation}
The solution approaching the Minkowski vacuum $e^{-ik\tau}/\sqrt{2k}$ in the far past is given in terms of a Hankel function of the first kind,
\begin{equation}
 f_k(\tau) = \sqrt{\frac{\pi}{4k}}e^{i(2\alpha_\sigma+1)\pi/4} \sqrt{-k\tau} H^{(1)}_{\alpha_\sigma} (-k\tau) \,.
\end{equation}
The phase is immaterial in what follows, and hence we will drop it.

The explicit solution to the mode functions allows us to evaluate the correlator. Using the formula $\frac{\dd}{\dd x}H^{(1)}_\alpha(x) = H^{(1)}_{\alpha - 1}(x) - \frac{\alpha}{x}H^{(1)}_\alpha (x)$ we obtain the compact expression
\begin{equation} \label{eq:esinglefieldcorrelator}
 A_{\sigma\sigma} = i\left(\frac{\alpha_\sigma - 1/2}{\tau}+\frac{z'}{z}\right) +  ik \left(\frac{H^{(1)}_{{\alpha_\sigma} - 1}(-k\tau)}{H^{(1)}_{\alpha_\sigma}(-k\tau)}\right)^* = ik\left(\frac{H^{(1)}_{{\alpha_\sigma} - 1}(-k\tau)}{H^{(1)}_{\alpha_\sigma}(-k\tau)}\right)^*\,.
\end{equation}
Note that the term involving $z'/z$ has disappeared from $A_{\sigma\sigma}$. This cancellation follows directly from the definition $\alpha_\sigma^2=1/4+\tau^2z''/z$, which implies that $z$ behaves as a power law $z \propto (-\tau)^n$ for some real $n$, in turn implying that $\alpha_\sigma^2 = (\tau z'/z-1/2)^2$. Thus for any single field model, where the fluctuations are adiabatic, i.e., those of the field that drives the background, the term inversely proportional to the comoving horizon, $z'/z\propto(-\tau)^{-1}\propto|\mathcal{H}|$, disappears from the correlator. This is important for models where the comoving horizon shrinks, such as inflation and ekpyrosis, since a potential source of amplification of perturbations is removed.  As we will see, the consequences of this fact for inflation and ekpyrosis differ drastically.

At early times ($\tau \to - \infty$), the correlator approximates its Minkoswki vacuum value $A_{\sigma \sigma} \simeq k$. In order to obtain the late-time limit, $\tau \to 0^-,$ one has to use the asymptotic expansion of the Hankel function,
\begin{align}
 H^{(1)}_\alpha(x) =& \left( \frac{x}{2}\right)^\alpha \left( \frac{1}{\Gamma(\alpha+1)} - \frac{x^2}{4\Gamma(\alpha+2)}+ \ldots \right) \nonumber \\
 & + i \left( \frac{x}{2}\right)^{-\alpha} \left( -\frac{\Gamma(\alpha)}{\pi} - \frac{x^2 \Gamma(\alpha-1)}{4\pi}+ \ldots - \left( \frac{x}{2}\right)^{2\alpha} \frac{\cos(\pi \alpha)\Gamma(-\alpha)}{\pi} + \ldots \right)\,. \label{Hankel}
\end{align}
The leading real and imaginary parts of the correlator depend on the value of $\alpha_\sigma$ and are given (in the late-time limit) by
\begin{subequations}
\begin{align}
 A_{\sigma\sigma} &\simeq\frac{k}{2 ^{ 2 \alpha_\sigma-1} \Gamma( \alpha_\sigma)}\left(\frac{\pi}{\Gamma( \alpha_\sigma)} - i \, \Gamma(1 - \alpha_\sigma)\cos( \pi \alpha_\sigma)\right)(-k\tau)^{2\alpha_\sigma-1} \,, \quad & (\alpha_\sigma < 1) \label{eq:corralphasmall} \\
 A_{\sigma\sigma} &\simeq k\left(\frac{\pi}{ 2 ^{ 2 \alpha_\sigma-1} \Gamma( \alpha_\sigma)^2} \, (-k\tau)^{2 \alpha_\sigma - 1} + i\,\frac{1}{2(\alpha_\sigma-1)} \, (-k\tau)\right)\,. \quad & (\alpha_\sigma > 1) \label{eq:corralphalarge}
\end{align}
\end{subequations}
Given that $z^2 = 2 \epsilon \bar{a}_0^2 |\tau|^{1-2\alpha_\sigma}$ when $0\leq\epsilon<1$ or $\epsilon\geq 3$, we can also find the corresponding expressions for the correlator for the comoving curvature perturbation $\mathcal{R}$,
\begin{subequations}
\begin{align}
 A_{\mathcal{R}\mathcal{R}} &\simeq\frac{\epsilon \bar{a}_0^2k^{2 \alpha_\sigma}}{2^{2\alpha_\sigma}\Gamma(\alpha_\sigma)}\left(\frac{\pi}{\Gamma( \alpha_\sigma)} - i \, \Gamma(1 - \alpha_\sigma)\cos( \pi \alpha_\sigma)\right)\,, \quad & (\alpha_\sigma < 1) \label{eq:corrZetaalphasmall} \\
 A_{\mathcal{R}\mathcal{R}} &\simeq \epsilon \bar{a}_0^2k^{2 \alpha_\sigma}\left(\frac{\pi}{ 2 ^{ 2 \alpha_\sigma} \Gamma( \alpha_\sigma)^2} \, + i \, \frac{1}{(\alpha_\sigma-1)} \, (-k\tau)^{2-2\alpha_\sigma}\right)\,. \quad & (\alpha_\sigma > 1) \label{eq:corrZetaalphalarge}
\end{align}
\end{subequations}
We will analyse the implications of these expressions for relevant single field models of the early universe in the next sections.

Given that during inflation and ekpyrosis the wave function evolves a lot, we can expect the complexity to grow significantly, too. In previous works, the complexity\footnote{Previous works used the analytically continued formula \eqref{eq:defcomplexity} for the definition of complexity \cite{Bhattacharyya:2020rpy,Bhattacharyya:2020kgu}. As already mentioned, we will rather use the hyperbolic measure.} was analysed with regard to the canonically normalised variable $v$ \cite{Bhattacharyya:2020rpy,Bhattacharyya:2020kgu}. In that case it is natural to use the early Minkowski correlator $A_{\sigma\sigma} \simeq k$ as the reference state. The target state is taken to be the late-time super-horizon state in which the perturbations find themselves at the end of inflation or at the end of the ekpyrotic scenario, just before reheating occurs in every case. Based on the definition \eqref{eq:defcomplexity2} --- so using the hyperbolic measure rather than the analytically continued one --- we may thus define the complexity of the canonical variable $v$ as\footnote{Here and below in Eq.~\eqref{eq:compR} we are summing over modes with wavenumbers $\mathbf{k}$ and $-\mathbf{k}$, because momentum conservation implies that these are produced together. Doing so introduces an additional overall factor of $\sqrt{2}$ in the complexities in comparison with Eq.~\eqref{eq:defcomplexityCh2}.}
\begin{equation}
 \mathcal{C}_v = \frac{1}{\sqrt{2}}\ln\left(X_v + \sqrt{X_v^2 - 1}\right)\,,\qquad X_v=\frac{1}{2}\left(\frac{\Re A_{\sigma\sigma}}{k}+\frac{k}{\Re A_{\sigma\sigma}}+\frac{(\Im A_{\sigma\sigma})^2}{k\Re A_{\sigma\sigma}}\right)\,.\label{eq:compv}
\end{equation}
As we will see below, it may make more sense, however, to work in terms of the physical variable, which is the comoving curvature perturbation. Then the complexity must be defined with respect to the state at some early time, which one may think of as the start of inflation or ekpyrosis. We will take the initial time to be $\tau_\mathrm{i}$ and the final time at reheating $\tau_\mathrm{f}$. Implicit in this prescription is the assumption that inflation and ekpyrosis had a start and did not reach back arbitrarily far into the past. This assumption seems well justified in light of recent works analysing the beginning stages of inflation \cite{East:2015ggf,Clough:2016ymm,Clough:2017efm,Aurrekoetxea:2019fhr,Hofmann:2019dqu,DiTucci:2019xcr,Bedroya:2019snp,Bedroya:2019tba,Brandenberger:2019eni,Joana:2020rxm,Jonas:2021xkx},
and it is well founded for ekpyrosis,
in particular in the context of its cyclic realisations \cite{Steinhardt:2001st,Steinhardt:2002ih,Khoury:2003rt,Steinhardt:2004gk,Lehners:2008qe,Lehners:2009eg,Ijjas:2019pyf}.
Starting again from the definition in Eq.~\eqref{eq:defcomplexity2}, we may thus write the complexity of the curvature perturbation as
\begin{align}
 \mathcal{C}_\mathcal{R}&=\frac{1}{\sqrt{2}}\ln\left(X_\mathcal{R}+\sqrt{X_\mathcal{R}^2-1}\right)\,,\nonumber\\
 X_\mathcal{R}(\tau)&=\frac{1}{2}\left(\frac{\Re A_{\mathcal{R}\mathcal{R}}(\tau)}{A_{\mathcal{R}\mathcal{R}}(\tau_\mathrm{i})}+\frac{A_{\mathcal{R}\mathcal{R}}(\tau_\mathrm{i})}{\Re A_{\mathcal{R}\mathcal{R}}(\tau)}+\frac{[\Im A_{\mathcal{R}\mathcal{R}}(\tau)]^2}{A_{\mathcal{R}\mathcal{R}}(\tau_\mathrm{i})\Re A_{\mathcal{R}\mathcal{R}}(\tau)}\right)\,,\label{eq:compR}
\end{align}
in the time interval $\tau_\mathrm{i}\leq\tau\leq\tau_\mathrm{f}$. This provides the basic ingredients to analyse the evolution of complexity in inflation and ekpyrosis in the following sections.

A few observations can immediately be made: as long as one remains deeply sub-horizon ($-k\tau\gg 1$), we notice that $A_{\sigma\sigma}\simeq k$ implies $X_v\simeq 1$ and $\mathcal{C}_v\simeq 0$, so complexity in terms of the canonical variable does not grow.
In terms of the curvature perturbation, the corresponding early-time sub-horizon limit is $A_{\mathcal{R}\mathcal{R}}(\tau)\simeq 2\epsilon ka(\tau)^2$, and so one has
\begin{equation}
 X_\mathcal{R}\simeq\frac{1}{2}\Big[\Big(\frac{a}{a_\mathrm{i}}\Big)^2+\Big(\frac{a_\mathrm{i}}{a}\Big)^2\Big]\,,\qquad(-k\tau\to\infty)
\end{equation}
where $a_\mathrm{i}\equiv a(\tau_\mathrm{i})$.
As the universe expands ($a\gg a_\mathrm{i}$) or contracts ($a\ll a_\mathrm{i}$), either one of the two terms will dominate.
Thus, as $X_\mathcal{R}\gg 1$, the complexity may be approximated as $\mathcal{C}_\mathcal{R}\simeq 2^{-1/2}\ln(2X_\mathcal{R})$, and so one finds
\begin{equation}
 \mathcal{C}_\mathcal{R}\simeq\pm\sqrt{2}\ln\left(\frac{a}{a_\mathrm{i}}\right)=\frac{\sqrt{2}}{|1-\epsilon|}\mathcal{N}\,,\label{eq:CRsubHgen}
\end{equation}
where the $+$ sign holds for expansion (when $0\leq\epsilon<1$) and the $-$ sign for contraction (when $\epsilon>1$).
Therefore on sub-horizon scales, as we will confirm in the next sections, inflation with $\epsilon\ll 1$ has $\mathcal{C}_\mathcal{R}\simeq\sqrt{2}\mathcal{N}$ and ekpyrosis with $\epsilon\gg 3$ has $\mathcal{C}_\mathcal{R}\simeq\sqrt{2}\mathcal{N}/\epsilon$.

\section{Inflation} \label{sec:inflation}

Inflation is characterised by accelerated expansion; hence $0 \leq \epsilon < 1$, and consequently $\alpha_\sigma>3/2$. In the slow-roll limit, where $\epsilon \ll 1,$ we may approximate $\alpha_\sigma \approx 3/2+\epsilon$; we will illustrate our results in this limit. One can see from Eq.~\eqref{eq:corralphalarge} that the dispersion,
\begin{equation}
 \frac{1}{\Re(A_{\sigma\sigma})} \sim |\tau|^{-2-2\epsilon}\,, \qquad (\epsilon \ll 1)
\end{equation}
of the canonically normalised modes $v$ is growing as inflation proceeds, i.e., as $\tau \to 0^-$. This means that these modes are strongly amplified. Meanwhile, since the ratio 
\begin{equation}
 \frac{\Im(A_{\sigma\sigma})}{\Re(A_{\sigma\sigma})} \sim |\tau|^{-1-2\epsilon} \qquad (\epsilon \ll 1)
\end{equation}
is also growing fast, we can see that the wave function evolves into a highly squeezed state. 

\begin{figure}[t]
\centering
\includegraphics[width=0.5\textwidth]{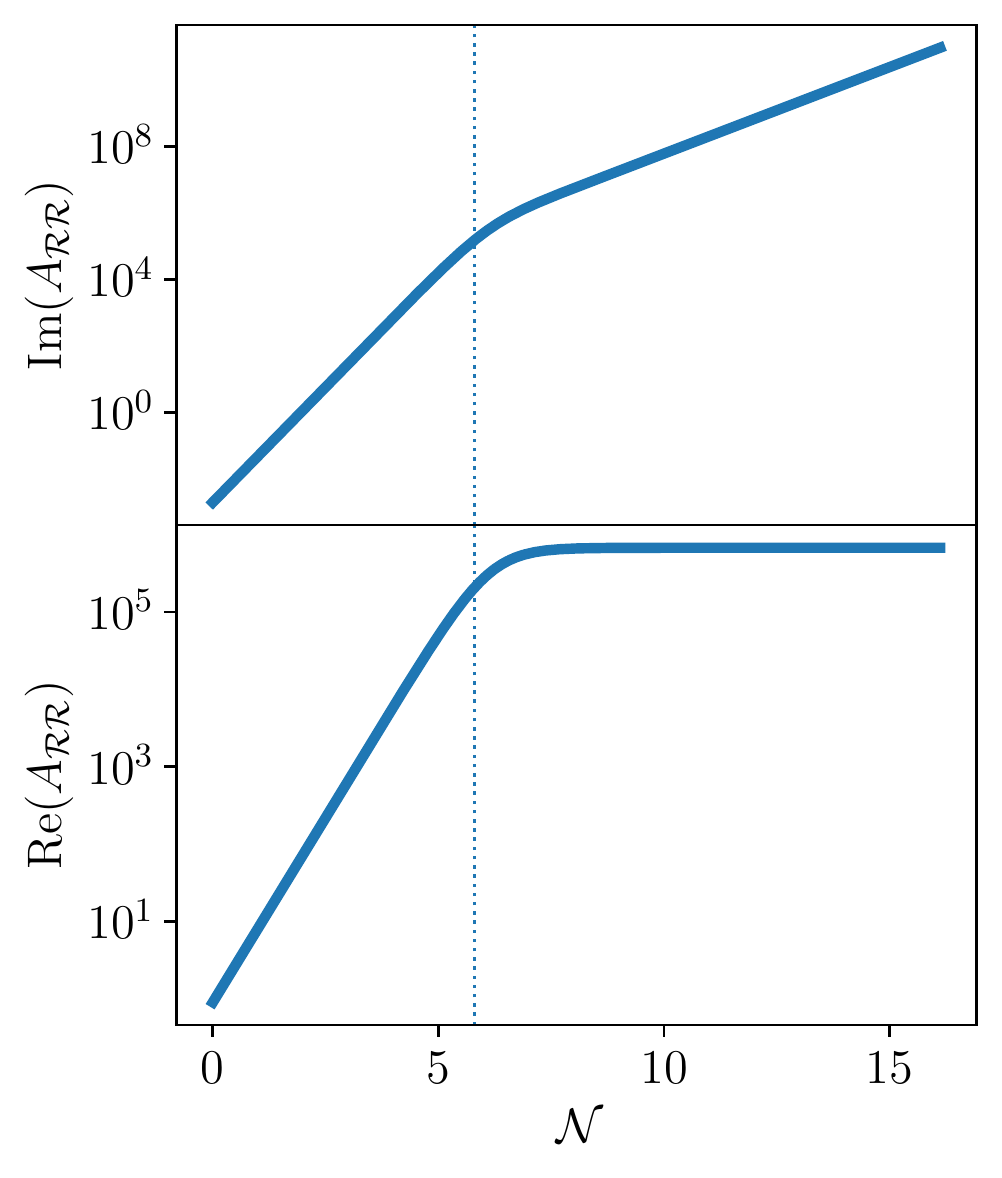}
\caption{Real (bottom) and imaginary (top) parts of the correlator for curvature perturbations $\mathcal{R}$ during inflation as a function of the $e$-folding number $\mathcal{N}$. The vertical, dotted line is the time of horizon exit. The following numerical values were used: $\epsilon = 1/11$ (so $\alpha_\sigma=8/5$) and $k=5$.} \label{fig:ARRInf}
\end{figure}

Another perspective is offered by the correlator for the physical perturbation, namely the comoving curvature perturbation, in Eq.~\eqref{eq:corrZetaalphalarge} --- see Fig.~\ref{fig:ARRInf}. From its late-time asymptotic expression, we can see that the real part evolves to a constant. This is a reflection of the fact that on large (super-horizon) scales, the comoving curvature perturbation becomes constant. The spectrum of the perturbations can be read off from the $k$ dependence, which implies a spectral index $n_\mathrm{s} = 4-2\alpha_\sigma \approx 1-2\epsilon$ with a small red tilt. Since the state is highly squeezed, $|\Im(A_{\sigma\sigma})/\Re(A_{\sigma\sigma})|\gg 1$, the fact that the curvature perturbation is approximately conserved on large scales implies that on these scales the momentum is known to be small with high precision and consequently the bulk of the uncertainty resides in the amplitude.

\begin{figure}[t]
\centering
\includegraphics[width=0.8\textwidth]{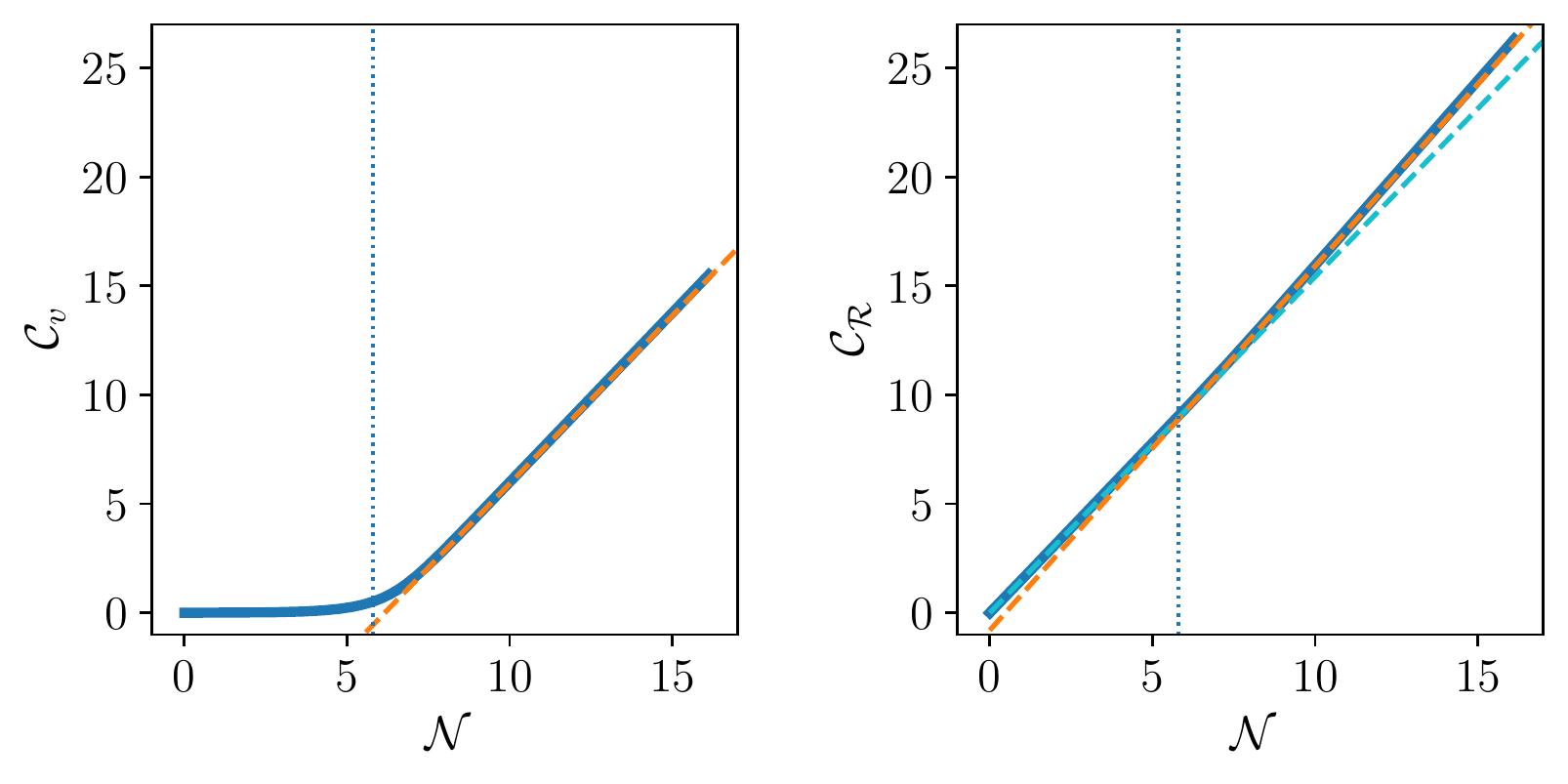}
\caption{Complexity during inflation in terms of the modes $v$ (left) and the curvature perturbations $\mathcal{R}$ (right) depicted by the blue curves. The vertical, dotted line is the time of horizon exit as in Fig.~\ref{fig:ARRInf}. The same numerical values as in Fig.~\ref{fig:ARRInf} are also used. The cyan and orange dashed lines are asymptotic expressions before and after horizon exit, respectively [see Eq.~\eqref{eq:CinfasymN}].} \label{fig:CInf}
\end{figure}

Using the full expression for the correlator [Eq.~\eqref{eq:esinglefieldcorrelator}], one can compute the complexity given by either \eqref{eq:compv} or \eqref{eq:compR}.
A numerical example of the evolution of the two complexities, $\mathcal{C}_v$ for the mode functions and $\mathcal{C}_\mathcal{R}$ for the curvature perturbations, is shown in Fig.~\ref{fig:CInf}. It is striking how differently they evolve: the most obvious difference is that $\mathcal{C}_v$ only grows on super-horizon scales, while $\mathcal{C}_\mathcal{R}$ grows throughout. In fact, for $\mathcal{C}_v$ we have the following leading-order approximate expressions for modes that have not yet exited the horizon and for those that already have [using Eq.~\eqref{eq:corralphalarge} and Taylor expanding Eq.~\eqref{eq:compv} as $\tau\to 0^-$],
\begin{subequations}\label{eq:Cvinfasym}
\begin{align}
 \mathcal{C}_v(\tau) &\simeq 0\,, \quad & (\textrm{sub-horizon}) \label{eq:CvinfasymsubH} \\
 \mathcal{C}_v(\tau) &\simeq\sqrt{2}(1+\epsilon)\ln\Big(\frac{\tau_\star}{\tau}\Big)\,, \quad & (\textrm{super-horizon})
\end{align}
\end{subequations}
where $\tau_\star$ indicates the conformal time at horizon exit,
while the complexity of the curvature perturbation is well approximated by
\begin{subequations}\label{eq:CRinfasym}
\begin{align}
 \mathcal{C}_\mathcal{R}(\tau) &\simeq \sqrt{2}\ln\Big(\frac{a(\tau)}{a_\mathrm{i}}\Big)\,, \quad & (\textrm{sub-horizon}) \label{eq:CRinfasymsubH} \\
 \mathcal{C}_\mathcal{R}(\tau) &\simeq \mathcal{C}_\mathcal{R}(\tau_\star)+\sqrt{2}(1+2\epsilon)\ln\Big(\frac{\tau_\star}{\tau}\Big)\,. \quad & (\textrm{super-horizon}) \label{eq:CRinfasymsupH}
\end{align}
\end{subequations}
The above expression on sub-horizon scales was already derived in Eq.~\eqref{eq:CRsubHgen}, while for super-horizon scales we used Eq.~\eqref{eq:corrZetaalphalarge} and expanded Eq.~\eqref{eq:compR} as $\tau\to 0^-$.
These expressions can be recast as the growth of complexity ($\Delta\mathcal{C}$) in the different regimes as a function of the duration of that regime ($\Delta\mathcal{N}$) as
\begin{subequations}\label{eq:CinfasymN}
\begin{align}
 \Delta\mathcal{C}_v &\simeq 0\,, \quad & (\textrm{sub-horizon}) \label{eq:CvinfasymsubHN} \\
 \Delta\mathcal{C}_v &\simeq \sqrt{2}(1+\epsilon)\Delta\mathcal{N}\,, \quad & (\textrm{super-horizon})\\
 \Delta\mathcal{C}_\mathcal{R} &\simeq\frac{\sqrt{2}}{1-\epsilon}\Delta\mathcal{N}\simeq\sqrt{2}(1+\epsilon)\Delta\mathcal{N}\,, \quad & (\textrm{sub-horizon}) \label{eq:CRinfasymsubHN} \\
 \Delta\mathcal{C}_\mathcal{R} &\simeq \sqrt{2}(1+2\epsilon)\Delta\mathcal{N}\,. \quad & (\textrm{super-horizon}) \label{eq:CRinfasymsupHN}
\end{align}
\end{subequations}
These approximate expressions are very accurate, as may be seen by the closeness of the numerical curve to the dashed lines in Fig.~\ref{fig:CInf}. The `canonical' complexity grows roughly as $\sqrt{2}$ times the number of $e$-folds that a given mode spends outside of the horizon. For modes that exit the horizon say $50$ $e$-folds before the end of inflation, the complexity will thus end up being approximately $50\sqrt{2}$. Meanwhile, modes that leave just before the end of inflation would have essentially vanishing complexity by the time of reheating. By contrast, the ``curvature complexity'' already grows (again roughly as $\sqrt{2}\Delta\mathcal{N}$) while a mode is still sub-horizon. This is because the curvature perturbation is significantly amplified as it is drawn out of the vacuum. Once the perturbation exits the horizon, the complexity keeps growing (roughly at the same rate) due to the imaginary part of the correlator, which is still growing. Thus, even though the real part of the correlator becomes constant, signalling that the curvature perturbation is conserved outside of the horizon, the wave function has a rapidly growing imaginary part. In other words, the wave function for the curvature perturbation is increasingly of Wentzel-Kramers-Brillouin (WKB) form, with a rapidly changing phase and a slowly changing amplitude. The phase drives the complexity to even higher values. Overall, the curvature complexity is thus much higher than the canonical complexity. If an inflationary phase lasts say $80$ $e$-folds, then the same mode considered above would undergo a growth of complexity of about $30\sqrt{2}\approx 42$ before horizon exit, followed by an additional $50\sqrt{2}\approx 71$ units of growth after horizon exit. More generally, if the inflationary phase lasts $\mathcal{N}_\mathrm{tot}$ $e$-folds in total, then the complexity of a mode exiting $\mathcal{N}_\star$ $e$-folds before the end is given by
\begin{equation}
 \mathcal{C}_v \simeq \sqrt{2}(1+\epsilon)\mathcal{N}_\star\,, \qquad \mathcal{C}_\mathcal{R} \simeq \frac{\sqrt{2}}{1-\epsilon}(\mathcal{N}_\mathrm{tot} - \mathcal{N}_\star) + \sqrt{2}(1+2\epsilon)\mathcal{N}_\star \simeq \sqrt{2}\left[\mathcal{N}_\mathrm{tot}+\epsilon(\mathcal{N}_\star+\mathcal{N}_\mathrm{tot})\right]\,.
\end{equation}
The canonical complexity is thus ignorant about the beginning and duration of inflation. By contrast, the curvature complexity is sensitive to the total duration of inflation and hence also to the pre-horizon-exit evolution.

The above result for the curvature complexity in inflation admits an additional interesting interpretation.
Approximating inflation by de Sitter spacetime and thinking of de Sitter as a thermal system with temperature given by the inverse of the horizon radius (i.e., $T\sim H$), one can ask the question of how fast the system can ``scramble'' \cite{Sekino:2008he,Susskind:2011ap} perturbations over the Hubble horizon.
From the point of view of a dynamical system, trajectories moving apart exponentially fast as a function of time ($\sim e^{\lambda_\mathrm{L}t}$) can be diagnosed as chaotic, and the strength of the sensitivity to the initial conditions is characterised by the Lyapunov exponent $\lambda_\mathrm{L}$.
The time at which trajectories have moved apart by an $\mathcal{O}(1)$ factor is then representative of the scrambling time ($t_*$), hence $t_*\sim 1/\lambda_\mathrm{L}$.
For a thermal quantum system, it has been conjectured that there exists an upper bound on the growth of chaos given by the temperature, namely $\lambda_\mathrm{L}\lesssim T$ \cite{Maldacena:2015waa}, and black holes as well as de Sitter are potentially among the fastest scrambling systems in the universe, given that they saturate the conjectured bound \cite{Sekino:2008he,Susskind:2011ap}.
For generic quantum systems, there exist various approaches and techniques to quantify the chaoticity and correspondingly compute the Lyapunov exponent and scrambling time (for de Sitter we certainly expect $\lambda_\mathrm{L}\sim T\sim H$; see, e.g, \cite{Aalsma:2020aib,Geng:2020kxh,Haque:2020pmp}).
Complexity might be a promising quantity in that respect.
Indeed, as the evolution of complexity depends on the logarithm of the correlator, linear growth in complexity is actually indicative of exponential separation of trajectories in ``field space'' (the space of quantum states) and thus of chaos \cite{Ali:2019zcj,Bhattacharyya:2019txx,Yan:2020wkt,Yan:2020twr,Bhattacharyya:2020art,Bhattacharyya:2020iic}.
One could therefore attempt to identify $\lambda_\mathrm{L}\equiv\mathrm{d}\mathcal{C}/\mathrm{d}t,$ which would yield a scrambling time of order the Hubble parameter, $\lambda_\mathrm{L}\simeq\sqrt{2}H$ (on super-horizon scales for $\mathcal{C}_v$, but on all scales for $\mathcal{C}_\mathcal{R}$). There might, however, be different roles for the canonical and curvature complexities: the canonical complexity may be a good measure of chaos in the sense that it is the effective mass squared of the mode functions that transitions from positive to negative near horizon exit, implying an enhanced sensitivity to initial conditions as the horizon is crossed. On the other hand, the physical complexity is sensitive to the entire inflationary evolution, and thus to the vast separation of scales achieved over the course of an entire inflationary phase. This may also lead to connections with the trans-Planckian censorship conjecture \cite{Bedroya:2019tba,Bedroya:2019snp,Brandenberger:2019eni,Bedroya:2020rmd}. It will certainly be of interest to explore the connections between complexity and chaos in much more detail in the future.

\section{Single-field ekpyrosis} \label{sec:ekpyrosis}

Ekpyrosis is a phase of slow contraction with ultra-stiff EoS $\epsilon > 3$. This EoS is required in order to make the ekpyrotic phase an attractor, such that anisotropies do not grow despite the fact that the universe is contracting. During this phase, the scale factor is almost constant, but the Hubble rate grows quickly in magnitude. Since the scale factor evolves little, the wavelength of perturbations changes equally little. However, the rapidly growing Hubble rate implies that the comoving horizon $1/|aH| \propto (-\tau)$ shrinks rapidly as the universe contracts, $\tau \to 0^-$. Consequently perturbation modes with ever shorter wavelengths successively exit the horizon, just as in inflation.

\begin{figure}[t]
\centering
\includegraphics[width=0.5\textwidth]{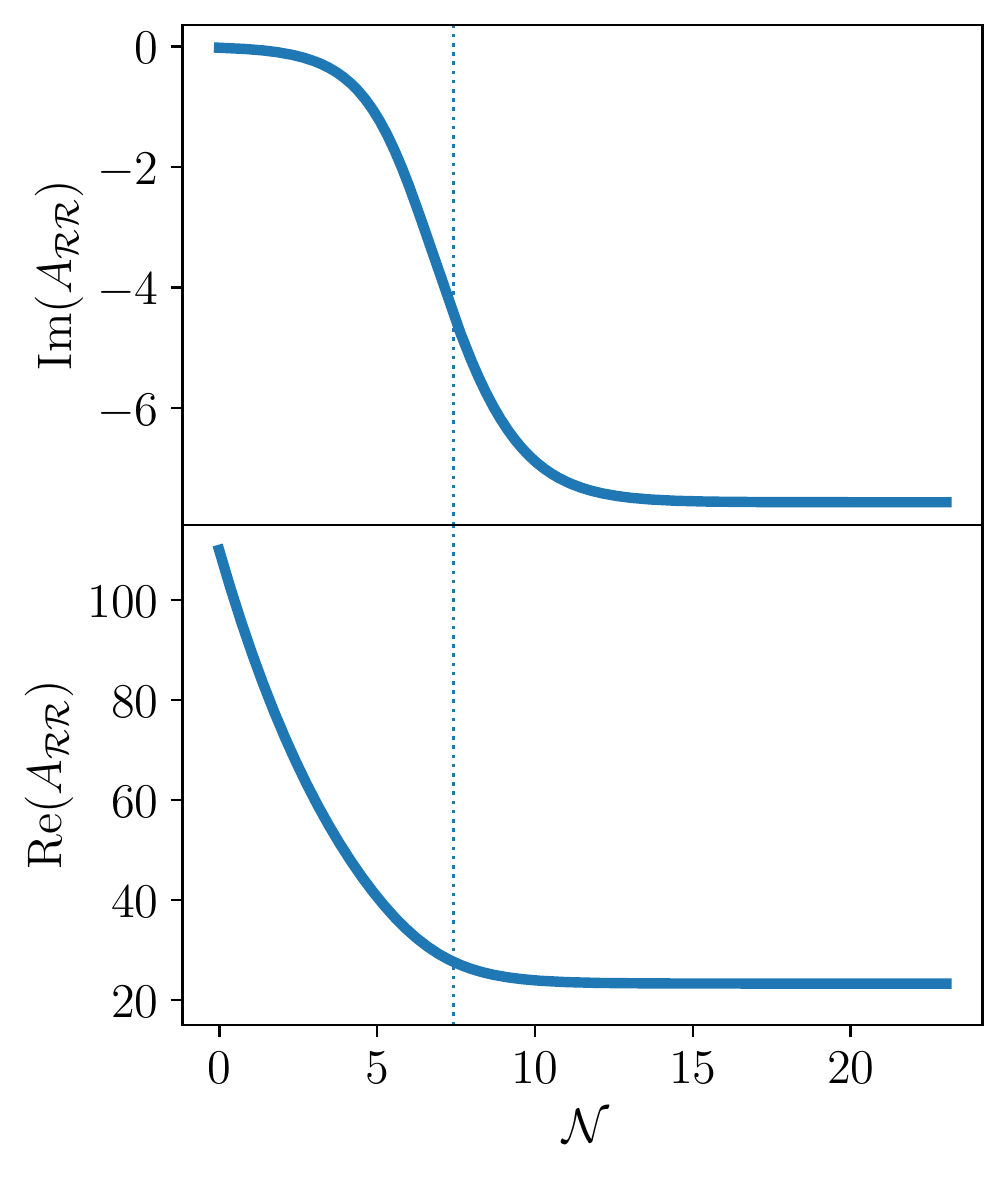}
\caption{A numerical evaluation of the correlator for the curvature perturbation $\mathcal{R}$ during ekpyrosis, with $k=5,$ $\epsilon = 11$, and consequently $\alpha_\sigma=2/5$. The vertical, dotted line indicates horizon exit. At late times the correlator tends to a constant.} \label{fig:ARREk}
\end{figure}

A crucial difference with inflation is that the adiabatic modes associated with the ekpyrotic scalar field do not get amplified much and do not develop a nearly scale-invariant spectrum. One may see this by inspection of the correlator for the curvature perturbation. Since the EoS is ultra-stiff, we have that $\alpha_\sigma < 1/2$ (for large $\epsilon$ we have $\alpha_\sigma \approx 1/2 - 1/\epsilon$) and thus Eq.~\eqref{eq:corrZetaalphasmall} implies that at late times both real and imaginary parts of the correlator $A_{\mathcal{R}\mathcal{R}}$ tend to constants. A numerical example is shown in Fig.~\ref{fig:ARREk}. Thus the ratio of the imaginary part to the real part does not grow, implying that no squeezing occurs and that these perturbations cannot be given a stochastic classical interpretation. Moreover, the spectral index $n_\mathrm{s} = 4 - 2\alpha_\sigma >3$ remains very blue.

\begin{figure}[t]
\centering
\includegraphics[width=0.8\textwidth]{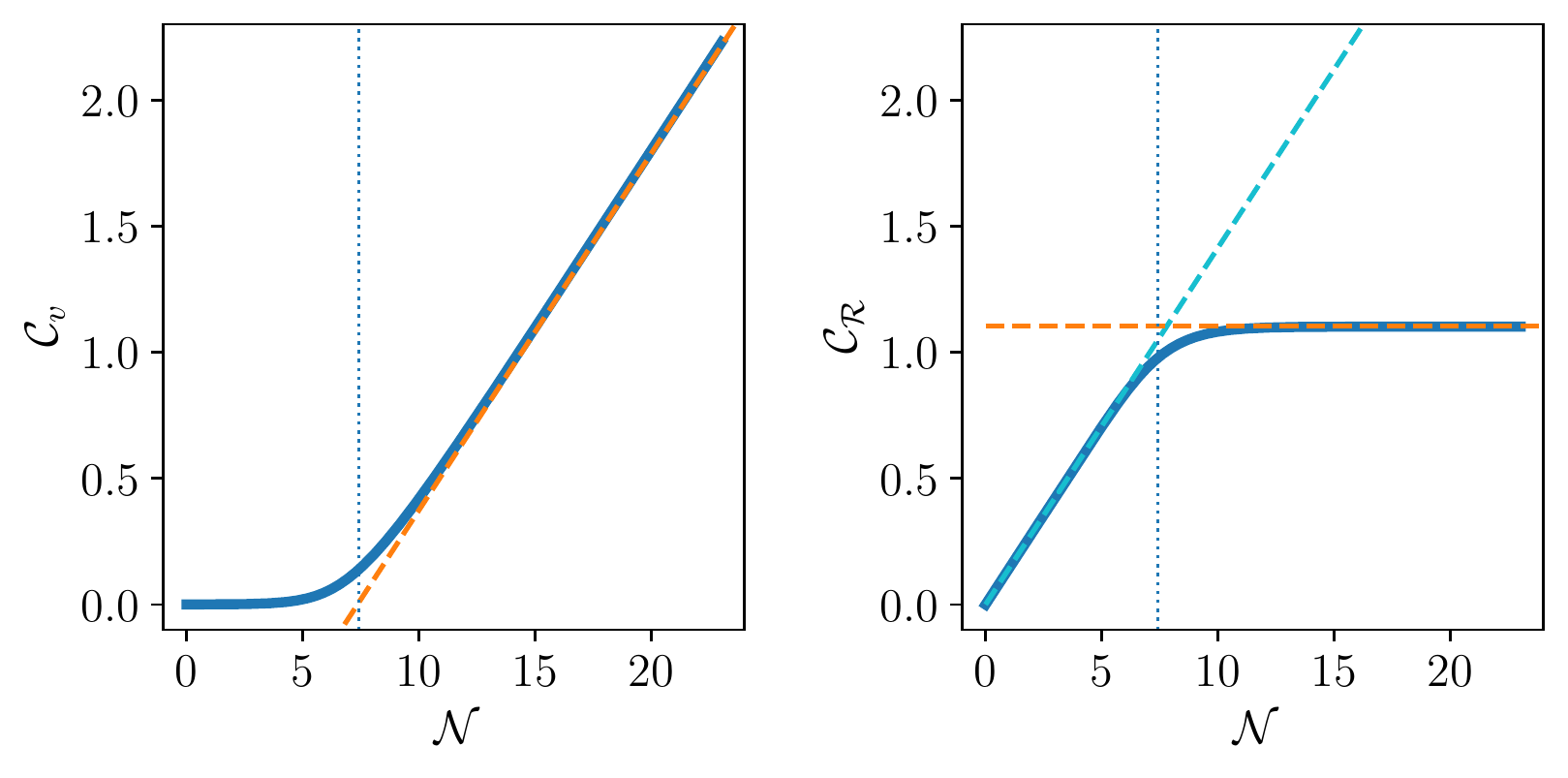}
\caption{The evolution of complexity during ekpyrosis, shown in terms of both the canonical complexity (left panel) and the curvature complexity (right panel), for the perturbation modes shown in Fig.~\ref{fig:ARREk}. The canonical complexity keeps growing even after horizon exit, while the curvature complexity reaches a constant value. As discussed in the main text, the curvature complexity seems to give a more faithful representation of the salient physical effects.} \label{fig:CEk}
\end{figure}

Given these observations, we may conclude that the wave function of adiabatic modes does not evolve much during an ekpyrotic phase, and thus we expect the complexity to remain rather small. A numerical example is shown in Fig.~\ref{fig:CEk}. The right panel of the figure shows the complexity of the curvature perturbation, which indeed reaches a small constant (depicted by the dashed orange line) soon after horizon exit --- before horizon exit it slowly grows as $\sqrt{2}\mathcal{N}/\epsilon$ (depicted by the dashed cyan line) as derived in Eq.~\eqref{eq:CRsubHgen}. By contrast, the canonical complexity (in the left panel) keeps growing even after horizon exit, due to the fact that the correlator for the Mukhanov-Sasaki variable keeps growing. Equation \eqref{eq:corralphasmall} shows that it grows in magnitude even at late times, with real and imaginary parts growing at the same rate as a result of the absence of squeezing [specifically $A_{\sigma\sigma}\sim(-\tau)^{-2/\epsilon}$, and so $\mathcal{C}_v\sim\sqrt{2}\mathcal{N}/\epsilon$, which is depicted by the dashed orange line in the left panel of the figure]. Our discussion in the previous paragraph indicates that the curvature complexity represents a more sensible reflection of the physical processes during the ekpyrotic phase, and thus constitutes the preferred measure of complexity (as was the case for inflation).

\section{Two-field ekpyrotic models} \label{sec:twofields}

In the previous section we saw that the adiabatic modes are not significantly amplified during an ekpyrotic phase, and moreover they retain a blue spectrum. Thus, put simply, they cannot explain the primordial temperature fluctuations seen in the cosmic background radiation and are utterly unimportant on large scales. However, the ekpyrotic background solution does possess a scaling symmetry, in close analogy to inflationary spacetimes. The problem, most clearly discussed in \cite{Creminelli:2004jg}, 
is that during ekpyrosis the scaling symmetry projects out of the curvature perturbation. A resolution of this issue is to consider two-field models. If there is a second ekpyrotic field, then the difference of the two fields (which inherits the scaling symmetry from both fields) is a gauge-invariant variable and may serve as the seed for primordial (nearly) scale-invariant perturbations. This has been discussed at length in several papers, e.g., \cite{Finelli:2002we,Notari:2002yc,DiMarco:2002eb,Lehners:2007ac,Koyama:2007mg,Koyama:2007ag}; hence we will only provide a brief review here. 

In addition to the adiabatic scalar $\sigma$ driving the background we will consider the transverse field $s$, whose fluctuations represent entropy (or isocurvature) perturbations. By definition we will assume that the ekpyrotic background evolution remains unchanged, so that at the background level $s=0$. (In the case where one starts with two ekpyrotic fields, this corresponds to rotating the field basis such that $\sigma$ points along the background trajectory.) The perturbations $\delta s$ will, however, be important. They are sensitive to the shape of the potential, which we take to be given by
\begin{equation}
 V(\sigma, s) = - V_0 e^{\sqrt{2\epsilon}\sigma}\left[ 1 + \frac{\kappa_2}{2} \epsilon s^2 +\mathcal{O}(s^3)\right]\,,
\end{equation}
where $\kappa_2$ is a positive parameter. In a theory of two scalar fields with two exact exponential potentials, the above potential arises after the above-mentioned field rotation and leads to $\kappa_2=1$. Thus $\kappa_2$ may be thought of as parameterising the deviation from exact exponential potentials. Note that the potential is unstable in the $s$ direction --- this crucial feature will be responsible for the amplification of $\delta s$ perturbations. It also leads to a sharpening of the issue of initial conditions, with possible implications investigated in \cite{Lehners:2008qe,Lehners:2009eg,Lehners:2011ig}. Non-Gaussian corrections are encoded in higher-order terms in the potential --- these lead to interesting observational consequences (see, e.g., \cite{Koyama:2007if,Lehners:2007wc,Lehners:2008my,Lehners:2009qu,Lehners:2009ja,Lehners:2010fy,Fertig:2016czu}), but will not be important for our present work.

There exists a second class of models, again with two scalar fields, but where instead of having a potential that is unstable in the transverse direction one considers a non-minimal coupling between the two scalar fields. A judicious choice of coupling can allow the adiabatic field to transfer its scaling symmetry to the second scalar field, without introducing instabilities in the background dynamics. These models, and their observational consequences, were explored in \cite{Qiu:2013eoa,Li:2013hga,Ijjas:2014fja}. They have the advantage that they lead to non-Gaussian signatures that are significantly smaller, yet within reach of near-future observations \cite{Fertig:2013kwa,Fertig:2015ola}. 

In all of these models, the ekpyrotic phase flattens the universe, suppresses anisotropies and amplifies entropy perturbations. In the adiabatic direction, the potential is approximately a negative exponential. This cannot grow indefinitely towards ever larger magnitudes, and one thus expects the potential to turn off at some point. This will mark the end of the ekpyrotic phase, and with the influence of a potential having disappeared, the evolution afterwards is that of a phase dominated by the kinetic energy of the adiabatic field. We will sometimes refer to this kinetic phase as ``kination''. Kination is usually envisaged to be a rather short phase, immediately followed by a bounce (and reheating) into the standard hot big bang expanding phase of the universe. The details of the bounce and reheating are highly model dependent (see, e.g., \cite{Takamizu:2004rq,Maeda:2008zz,Battefeld:2007st,Quintin:2014oea,Hipolito-Ricaldi:2016kqq}), 
just as for reheating after inflation. Nevertheless, the long wavelength modes we are interested in remain essentially unchanged during a non-singular bounce (see, e.g., \cite{Battarra:2014tga,Quintin:2015rta,Quintin:2019orx}). Thus we will end our analysis with the kinetic phase, the same way we ended the analysis before inflationary reheating in the previous section.

The analysis is greatly simplified by the realisation that the adiabatic/curvature and entropic/isocurvature perturbations remain decoupled during ekpyrosis and most (or all) of kination, even for models where the background scalar fields are coupled non-minimally \cite{Ijjas:2014fja}. The Lagrangian is thus a sum of terms involving solely the adiabatic mode $v_\sigma \equiv z \mathcal{R}$ and terms involving only the entropic mode $v_s \equiv a \, \delta s$ (where from here on we will consider a single Fourier mode and drop the subscript indicating the wavenumber to lighten the notation):
\begin{equation}
 \mathcal{L}^{(2)} =  \frac{1}{2} v _{\sigma} ^{ \prime 2} + \frac{1}{2} v _{s} ^{ \prime 2} -\frac{z'}{z}v_{\sigma}'v_\sigma - \frac{a'}{a}v_s' v_s  - \frac{1}{2} \left(k^2 - \frac{z^{\prime 2}}{z^2} \right) v _{\sigma} ^2  - \frac{1}{2} \left( k^2 - \frac{a^{\prime 2}}{a^2}+ a^2 \partial^2_sV \right) v _{s} ^2\,.
\end{equation}
The equations of motion are thus
\begin{equation}
 v_\sigma^{\prime\prime} + \left(k^2-\frac{z''}{z}\right) v_\sigma = 0\,, \qquad v_s^{\prime\prime} + \left(k^2-\frac{a''}{a} + a^2 \partial^2_sV\right) v_s = 0\,, \label{twofieldeom}
\end{equation}
and the canonical momenta are given by
\begin{equation}
 \pi_\sigma = v_\sigma^\prime - \frac{z'}{z}v_\sigma\,, \qquad \pi_s = v_s^\prime - \frac{a'}{a}v_s\,. 
\end{equation}
The fields can be quantised by following the usual procedure of promoting fields to operators,
\begin{equation}
\begin{pmatrix}
  \hat{v}_\sigma  \\ 
  \hat{v}_s 
\end{pmatrix} = 
\begin{pmatrix}
  f_\sigma & g_\sigma\\ 
  f_s & g_s
\end{pmatrix}
\begin{pmatrix}
   \hat{a} \\ 
  \hat{b} 
\end{pmatrix} + \mathrm{H.c.}\,,
\end{equation}
with $f_{\sigma,s}(\tau),\,g_{\sigma,s}(\tau)$ being complex, linearly independent solutions of \eqref{twofieldeom}. The conserved Wronskian combinations are now slightly more involved and read
\begin{subequations}
\begin{align}
 f_\sigma f_{\sigma}^{*\prime} + f_s f_s^{*\prime}  - \mathrm{c.c.} &= i\,, \label{eq:wronskian1}\\
 g_\sigma g_{\sigma}^{*\prime} + g_s g_s^{*\prime} - \mathrm{c.c.} &= i\,, \\
 f_\sigma \left(g_{\sigma}^{\prime}-\frac{z'}{z} g_{\sigma} \right) + f_s \left(g_s^{\prime} - \frac{a'}{a} g_s\right) - (f \leftrightarrow g) &= 0\,, \\
 f_\sigma \left(g_{\sigma}^{*\prime}-\frac{z'}{z} g_{\sigma}^* \right) + f_s \left(g_s^{*\prime} - \frac{a'}{a} g_s^*\right) - (f \leftrightarrow g) &= 0\,. \label{eq:wronskian4}
\end{align}
\end{subequations}
All of this is simply the two-field generalisation of the discussion in section \ref{sec:quantisation}. One can now use the Wronskians to express the annihilation operators in terms of the field operators and momenta,
\begin{subequations}
\begin{align}
 i\, \hat{a} & =  \left(f_{\sigma}^{*\prime}-\frac{z'}{z} f_{\sigma}^*\right) \hat{v}_\sigma -f_{\sigma}^* \hat{\pi}_\sigma +\left(f_s^{*\prime} - \frac{a'}{a} f_s^*\right) \hat{v}_s - f_s^* \hat{\pi}_s \;,\\
 i\,\hat{b} & =  \left(g_{\sigma}^{*\prime}-\frac{z'}{z} g_{\sigma}^*\right) \hat{v}_\sigma -g_{\sigma}^* \hat{\pi}_\sigma +\left(g_s^{*\prime} - \frac{a'}{a} g_s^*\right) \hat{v}_s - g_s^* \hat{\pi}_s \;.
\end{align}
\end{subequations}
This allows us to transition to the Schr\"{o}dinger picture, where the vacuum wave function is defined via $\hat{a}|\Psi\rangle = \hat{b}|\Psi\rangle = 0$.
Thus, with the commutation relations realised via $\hat\pi \to -i\partial_v$, and up to an unimportant normalisation proportionality factor, the wave function is given by
\begin{equation}
 |\Psi(v_\sigma,v_s)\rangle \propto \exp \left( - \frac{1}{2}A_{ \sigma \sigma} v_ \sigma ^2 - \frac{1}{2} A_{ s s} v_ s ^2 \right) \,, \label{eq:wave function2}
\end{equation}
with the correlators
\begin{subequations}
\begin{align}
 A_{ \sigma \sigma} & =  - i \frac{ g_{s} ^{*} f_{\sigma}^{*\prime}-f_s^* g_{\sigma}^{*\prime}}{ g_{s}^{*} f_{\sigma}^{*}  - f_{s} ^{*} g_{\sigma}^{*}} + i \frac{z'}{z} \,,\\ \label{eq:solCorr2}
 A_{ ss} & =  - i \frac{  f_{\sigma}^{*}g_{s}^{*\prime}  - g_{\sigma}^{*} f_{s}^{*\prime} }{ f_{\sigma}^{*} g_{s}^{*} - g_{\sigma}^{*} f_{s} ^{*} } + i \frac{a'}{a} \,.
\end{align}
\end{subequations}

During the ekpyrotic phase, the scale factor evolves as a function of conformal time $\tau$ the same way as in Eq.~\eqref{eq:backscal} and consequently 
\begin{equation}
  \frac{a''}{a}  =  \frac{z''}{z} = - \frac{ \epsilon -2}{( \epsilon -1) ^2} \frac{1}{ \tau ^2} \,, \qquad
  \frac{a''}{a} - a ^2\partial^2_sV  \approx \left(2\kappa_2 - \frac{2\kappa_2}{\epsilon} - \frac{1}{\epsilon}\right) \frac{1}{ \tau ^2}\,.
\end{equation}
The Fourier mode functions for the adiabatic ($v_\sigma$) and entropic ($v_s$) fields then have the solution
\begin{subequations}
\begin{align}
 f_{\sigma}(\tau) &= \sqrt{\frac{\pi}{4k}}\sqrt{-k\tau} H^{(1)}_{\alpha_{\sigma}}(-k\tau)\,, \quad f_s(\tau)=0\,,\\
 g_{s}(\tau) &= \sqrt{\frac{\pi}{4k}}\sqrt{-k\tau} H^{(1)}_{\alpha_{s}}(-k\tau)\,, \quad g_\sigma(\tau) = 0\,,
\end{align}
\end{subequations}
where the integration constants have been fixed such that at early times the mode functions approach their expressions in the Minkowski vacuum (up to an unimportant phase). The order of the Hankel functions is given by
\begin{equation}
 \alpha_\sigma = \frac{\epsilon - 3}{2(\epsilon - 1)}\,, \qquad \alpha_s \approx \left(\frac{1}{4}+2\kappa_2 - \frac{2\kappa_2}{\epsilon} - \frac{1}{\epsilon}\right)^{1/2}\,.
\end{equation}
There is no mixing between the modes. Hence for the adiabatic modes we will recover exactly what we had before in section \ref{sec:ekpyrosis}. Note that for $\kappa_2 \approx 1$ we have that $\alpha_s \approx 3/2$, and thus we can expect that by contrast the entropic modes will behave much more like inflationary perturbations. The correlators may be written as
\begin{subequations}
\begin{eqnarray} \label{eq:ekpyroticcorrelators}
 A_{ \sigma \sigma} & =  & ik \frac{H^{(1)*}_{{\alpha_\sigma} - 1}(-k\tau)}{H^{(1)*}_{\alpha_\sigma}(-k\tau)} \,,\\ 
 A_{ ss} & = &   -i\left[\left(\frac{1}{2}- \alpha_s\right)(\epsilon - 1)-1\right]\mathcal{H} + ik \frac{H^{(1)*}_{{\alpha_s} - 1}(-k\tau)}{H^{(1)*}_{\alpha_s}(-k\tau)} \,.
\end{eqnarray}
\end{subequations}
At early times, $\tau \to - \infty,$ the correlators approximate their Minkowski vacuum values $A_{\sigma \sigma} \simeq k$, $A_{ss} \simeq k$. At late times, we may again use the asymptotic form of the Hankel functions in Eq.~\eqref{Hankel} to obtain the same $A_{\sigma\sigma}$ as in Eq.~\eqref{eq:corralphasmall} and
\begin{equation}
 A_{ss}\simeq k\left[\frac{\pi}{2^{2\alpha_s-1}\Gamma(\alpha_s)^2}(-k\tau)^{2\alpha_s-1}-i\left(\frac{1}{2}-\alpha_s-\frac{1}{\epsilon-1}\right)(-k\tau)^{-1}\right]\,,\label{eq:corralphalarge2}
\end{equation}
where we have kept the leading real and imaginary parts. Given that the physical perturbation is $\delta s = v_s/a$, it makes more sense to consider its correlator $A_{\delta s \delta s} = a^2 A_{ss}$, which during the ekpyrotic phase evolves to
\begin{subequations}\label{approxentropycorr}
\begin{align}
 A_{\delta s\delta s}&\simeq \bar{a}_0^2\left[\frac{\pi k^{2\alpha_s}}{2^{2\alpha_s-1}\Gamma(\alpha_s)^2}(-\tau)^{2\alpha_s-\frac{\epsilon-3}{\epsilon-1}}-i\left(\frac{1}{2}-\alpha_s-\frac{1}{\epsilon-1}\right)(-\tau)^{-\frac{\epsilon-3}{\epsilon-1}}\right] \\
 &\approx\bar{a}_0^2\left[(-k\tau)^3+i\,\frac{\epsilon}{\epsilon-1}\right](-\tau)^{-\frac{\epsilon-3}{\epsilon-1}}\,. \qquad \qquad (\alpha_s \approx 3/2)
\end{align}
\end{subequations}
In the last line we made the approximation $\alpha_s \approx 3/2$. Several features can immediately be read off: since $\epsilon > 3$, the real part of the correlator shrinks, indicating that these modes will be amplified. Meanwhile, the imaginary part grows, the ratio between imaginary and real parts growing ever more rapidly as $|\tau|^{-3}$. Thus the state becomes highly squeezed and the perturbations become equivalent to a stochastic mixture of classical perturbations. In contrast to inflation, the dispersion of the entropy perturbations ($\sim 1/\Re[A_{\delta s\delta s}]$) does not reach a constant value, but rather keeps growing as the ekpyrotic phase proceeds. This is because the ekpyrotic potential is steep and keeps evolving to larger magnitudes. In order for the entropy perturbations to be able to act as realistic seeds for primordial density perturbations, they must reach a magnitude of around $10^{-4}$, implying that the potential must reach the grand unified scale \cite{Lehners:2007ac}. 

As discussed in section \ref{sec:ekpyrosis}, the potential turns off at this point and a kinetic phase ensues. For the purposes of our present study, it is a good approximation to assume that the potential becomes zero abruptly. We just have to make sure that we match the scale factor and the mode functions at the ekpyrosis-kination transition (call it $\tau_\textrm{e-k}$). This may be done as follows: during the kinetic phase, we may write the scale factor as $a_\mathrm{kin}(\tau)=a_{\mathrm{kin},0}(\tau_\mathrm{c} - \tau)^{1/2},$ allowing for the fact that the would-be crunch is shifted by an amount $\tau_\mathrm{c}$ compared to the ekpyrotic phase. Moreover, a general solution of the mode equation \eqref{twofieldeom} during the kinetic phase (with $V=0$) is given by a linear combination of Hankel functions of the first and second kinds as
\begin{equation}
 g_\mathrm{kin}(\tau) = \sqrt{-\tau}\left[ c_1 H^{(1)}_0 (-k\tau) + c_2 H^{(2)}_0 (-k\tau)\right]\,. \label{modekinetic}
\end{equation}
Then we can match the field values and momenta at the matching time $\tau_\textrm{e-k}$ by imposing
\begin{equation}
 a(\tau_\textrm{e-k})=a_\mathrm{kin}(\tau_\textrm{e-k})\,,\ \ a'(\tau_\textrm{e-k})=a_\mathrm{kin}'(\tau_\textrm{e-k})\,,\ \ g(\tau_\textrm{e-k})=g_\mathrm{kin}(\tau_\textrm{e-k})\,,\ \ g'(\tau_\textrm{e-k})=g_\mathrm{kin}'(\tau_\textrm{e-k})\,,
\end{equation}
and solving for the constants $a_{\mathrm{kin},0},\,\tau_\mathrm{c},\,c_1,\,c_2$.

\begin{figure}[t]
\begin{center}
\includegraphics[width=0.5\textwidth]{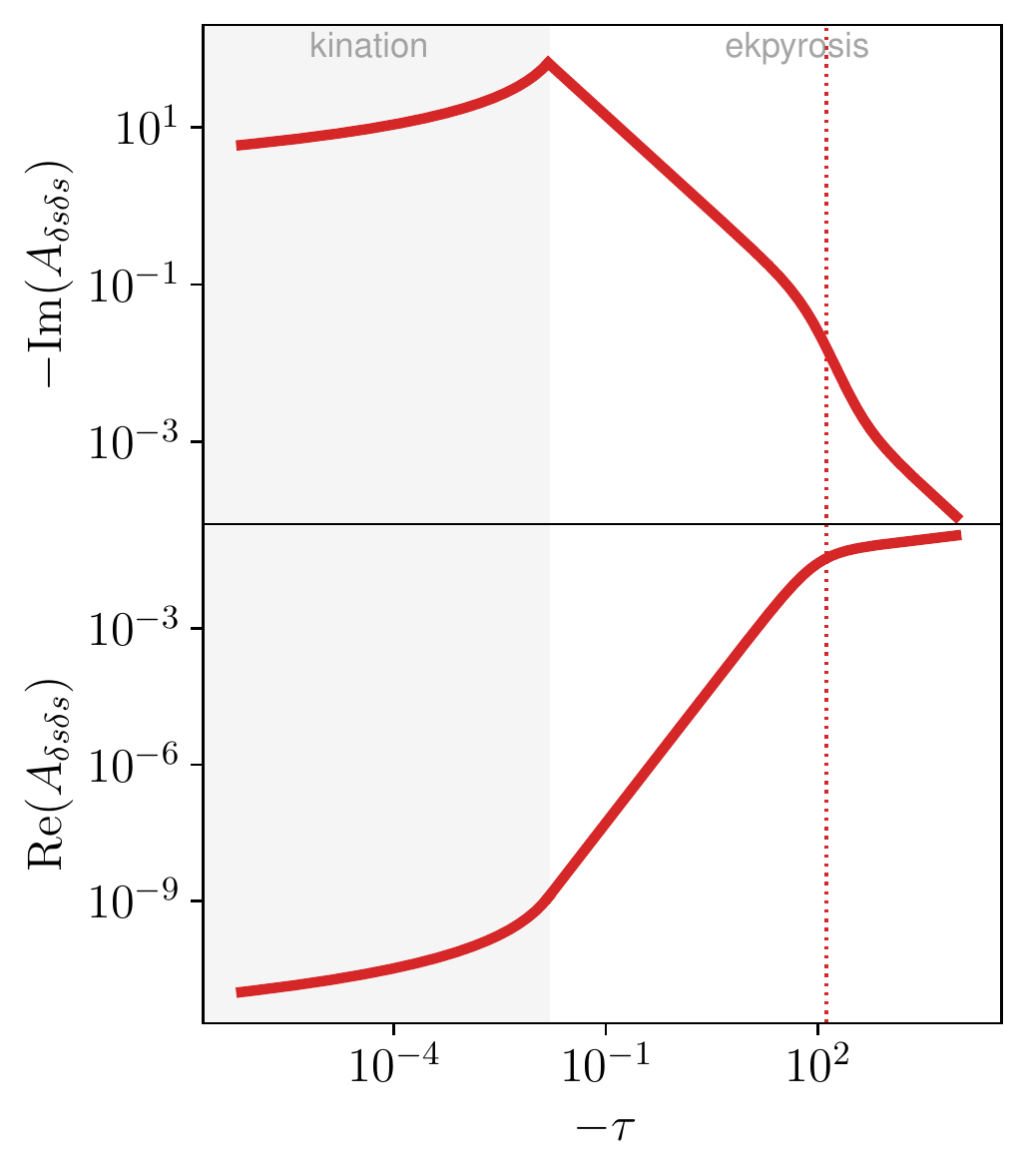}
\caption{The entropy correlator $A_{\delta s \delta s}=a^2A_{ss}$ during both ekpyrosis and kination as a function of conformal time for the following numerical values: $\kappa_2=1$, $\epsilon=12$, $k=0.01$. Time runs from right to left, and the vertical, dotted line depicts the time of horizon exit. The figure shows that the real part of the correlator is significantly reduced, implying that entropy perturbations are amplified. Meanwhile the imaginary part grows substantially (in absolute value), rendering the wave function highly squeezed. The kinetic phase reduces the real and imaginary parts of the correlator in tandem by a comparatively small amount, reflecting an additional modest growth of perturbations during kination.} \label{fig:twofieldcorrelator}
\end{center}
\end{figure}

The evolution of the correlator $A_{\delta s \delta s}$ is shown in Fig.~\ref{fig:twofieldcorrelator} as a function of conformal time. At the ekpyrotic-kinetic transition, the correlator is continuous, but its derivative is not. This is because the derivative of the correlator involves second derivatives of $a$ and $g$, and these are sensitive to the abrupt turning off of the potential. Figure \ref{fig:twofieldcorrelator} shows that during the kinetic phase, the entropy perturbations are further amplified since the real part of $A_{\delta s \delta s}$ keeps decreasing. This is, however, just a logarithmic growth, which does not alter the ekpyrotic amplification significantly. As for the imaginary part, it is similarly reduced during the kinetic phase, implying that the amount of squeezing remains approximately constant. We can verify this using a simple analytic approximation. In the large-scale limit, $-k\tau \ll 1,$ the mode function \eqref{modekinetic} during the kinetic phase can be approximated as $g_\mathrm{kin}(\tau)\simeq\sqrt{-\tau}[C_1+C_2\ln(-k\tau)]$, where the complex constants $C_{1,2}$ are related to $c_{1,2}$. At fixed $k$ and in the limit $\tau\to 0^-$, it is thus dominated by $g_\mathrm{kin}(\tau)\sim\sqrt{-\tau}\ln(-k\tau)$.
Then, the correlator for the entropy perturbation scales as (to leading order for the real and imaginary parts)
\begin{equation}
 A_{ss}\simeq\frac{\tilde{C}}{2(-\tau)[\ln(-k\tau)]^2}+\frac{i}{(-\tau)\ln(-k\tau)}\,,
\end{equation}
where $\tilde C$ is a real constant related to $C_{1,2}$ (and $c_{1,2}$ for that matter), and so for the physical perturbations, we have
\begin{equation}
 \Re(A_{\delta s\delta s})\sim\frac{1}{[\ln(-k\tau)]^2}\,,\qquad\Im(A_{\delta s\delta s})\sim\frac{1}{\ln(-k\tau)}\,.\label{eq:Adeltaskinapprox}
\end{equation}
The overall logarithmic correction is easily visible in Fig.~\ref{fig:twofieldcorrelator}.

\begin{figure}[t]
\begin{center}
\includegraphics[width=0.8\textwidth]{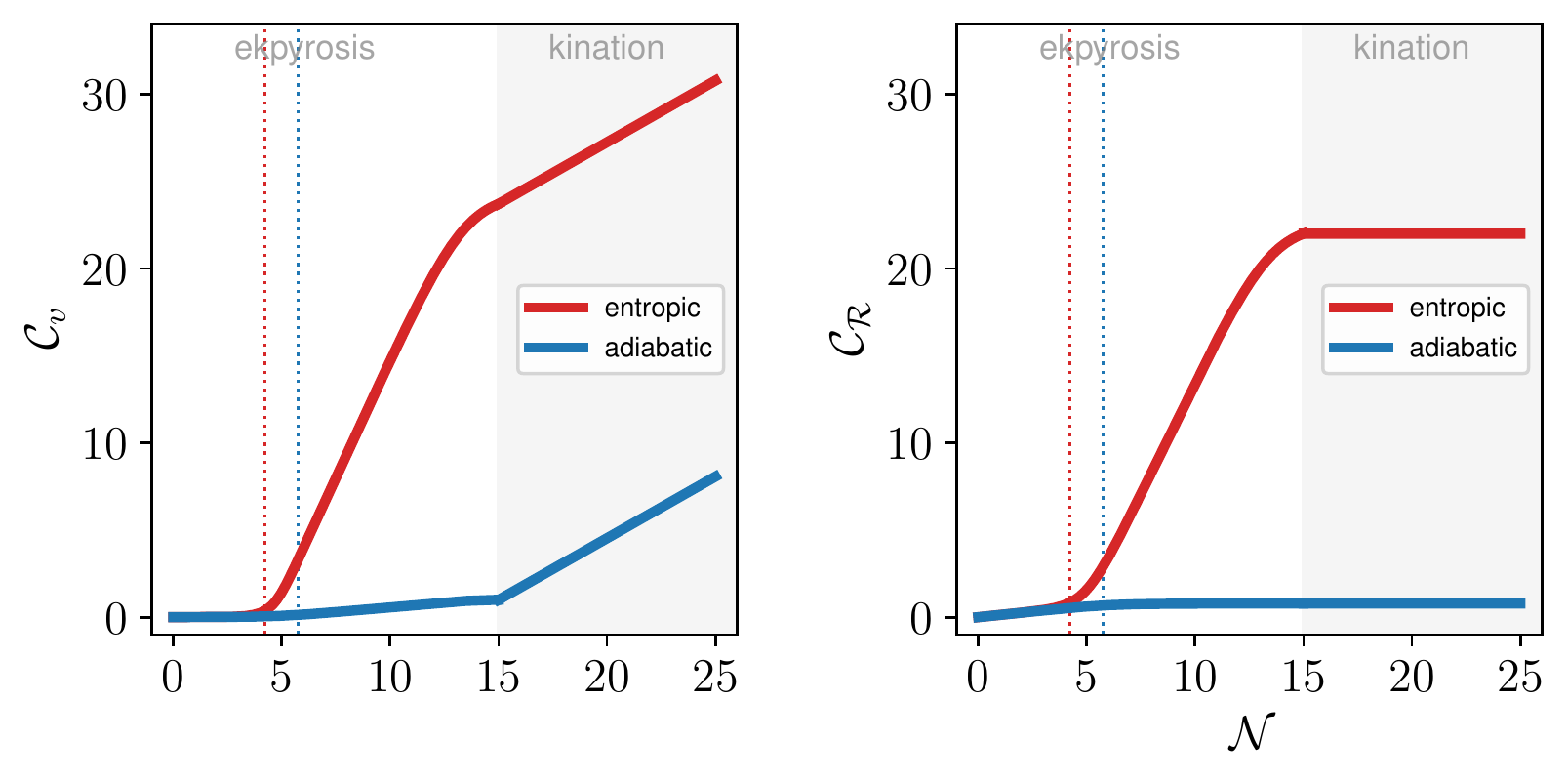}
\caption{The complexity of adiabatic (blue lines) and entropy (red lines) perturbations during ekpyrosis and kinetic domination as a function of the total $e$-folding number $\mathcal{N}\propto\ln(a|H|)$. In this example $k=0.01$ and $\epsilon = 12$ during the ekpyrotic phase, while $\epsilon = 3$ during kination. The vertical, dotted lines depict the time of horizon exit for the different modes, taken to be the time at which the effective mass squared of the mode functions turns negative. The left panel shows the complexity defined in terms of the mode functions (i.e., with respect to $f_\sigma$ and $g_s$ for adiabatic and entropy modes, respectively), which we argue in the text to be somewhat misleading. A better measure of complexity is that of the physical perturbations, shown in the right panel (so here $\mathcal{C}_\mathcal{R}$ denotes both $\mathcal{C}_\mathcal{R}$ and $\mathcal{C}_{\delta s}$, i.e., the complexity when taking the correlator to be $A_{\mathcal{R}\mathcal{R}}=z^2A_{\sigma\sigma}$ and $A_{\delta s\delta s}=a^2A_{ss}$ for adiabatic and entropy modes, respectively). The complexity of the adiabatic curvature perturbation remains very small, but that of the entropy perturbation is strongly enhanced. At a later stage, the entropy perturbation is envisaged to act as a source for the curvature perturbation, so that in the end the curvature perturbations will inherit the complexity of the entropy perturbations.} \label{fig:twofieldcomplexity}
\end{center}
\end{figure}

We are finally in a position to investigate the circuit complexity of the entropy perturbations. Here we also have the choice of looking at the complexity of the mode functions $v_s$ or that of the physical perturbation $\delta s$. Their respective expressions are given by
\begin{subequations}\label{eq:compsdeltas}
\begin{align}
 \mathcal{C}_{s,\delta s}(\tau) & = \frac{1}{\sqrt{2}}\ln\left(X_{s,\delta s}(\tau)+\sqrt{X_{s,\delta s}(\tau)^2-1}\right)\,,\\
 X_s(\tau)&=\frac{1}{2}\left(\frac{\Re A_{ss}(\tau)}{k}+\frac{k}{\Re A_{ss}(\tau)}+\frac{[\Im A_{ss}(\tau)]^2}{k\Re A_{ss}(\tau)}\right)\,,\\
 X_{\delta s}(\tau)&=\frac{1}{2}\left(\frac{\Re A_{\delta s\delta s}(\tau)}{A_{\delta s\delta s}(\tau_\mathrm{i})}+\frac{A_{\delta s\delta s}(\tau_\mathrm{i})}{\Re A_{\delta s\delta s}(\tau)}+\frac{[\Im A_{\delta s\delta s}(\tau)]^2}{A_{\delta s\delta s}(\tau_\mathrm{i})\Re A_{\delta s\delta s}(\tau)}\right)\,,
\end{align}
\end{subequations}
where $\tau_\mathrm{i}$ marks the start of the ekpyrotic phase. A numerical example showing the evolution of complexity is shown in Fig.~\ref{fig:twofieldcomplexity} for $\epsilon = 12$ and $k=0.01$. The figure shows both adiabatic and entropy perturbations, during ekpyrosis and kination, for both definitions of complexity. The adiabatic perturbations were discussed in section \ref{sec:ekpyrosis}, here the only novelty is that the evolution is followed into the kinetic phase. From the figure we can see that the canonical complexity grows significantly during the kinetic phase, even though the adiabatic modes evolve little, are not amplified and remain unsqueezed. This is again a reason to prefer the definition of complexity in terms of the physical variable, which remains essentially constant for the adiabatic modes throughout ekpyrosis and kination. Our real interest lies with the entropy perturbations. For these, the physical complexity grows significantly during the ekpyrotic phase, while being constant during the kinetic phase. With the aforementioned analytic series expansion for the correlators, we may find the following approximations for the complexity:
\begin{subequations}
\begin{align}
 \mathcal{C}_{\delta s}(\tau) & \simeq \sqrt{2}\ln\Big(\frac{a_\mathrm{i}}{a(\tau)}\Big)\,, &&\textrm{(sub-horizon)} \\
 \mathcal{C}_{\delta s}(\tau) & \simeq \mathcal{C}_{\delta s}(\tau_\star)+\sqrt{2}\left(\frac{2\epsilon-3}{\epsilon-1}\right)\ln\Big(\frac{\tau_\star}{\tau}\Big)\,, &&\textrm{(super-horizon, ekpyrotic)} \label{eq:CdelltassupH} \\
 \mathcal{C}_{\delta s}(\tau) & \simeq \mathcal{C}_{\delta s}(\tau_\textrm{e-k})=\mathrm{const.}\,, &&\textrm{(super-horizon, kination)}
\end{align}
\end{subequations}
and hence the complexity growth can be written as
\begin{subequations}
\begin{align}
 \Delta\mathcal{C}_{\delta s}(\tau) & \simeq \frac{\sqrt{2}}{\epsilon-1}\Delta\mathcal{N}\simeq\frac{\sqrt{2}}{\epsilon}\Delta\mathcal{N}\,, &&\textrm{(sub-horizon)} \\
 \Delta\mathcal{C}_{\delta s}(\tau) & \simeq \sqrt{2}\left(\frac{2\epsilon-3}{\epsilon-1}\right)\Delta\mathcal{N}\simeq 2\sqrt{2}\Delta\mathcal{N}\,, &&\textrm{(super-horizon, ekpyrotic)} \label{eq:CdelltassupHN} \\
 \Delta\mathcal{C}_{\delta s}(\tau) & \simeq 0\,, &&\textrm{(super-horizon, kination)} \label{eq:CdelltassupHK}
\end{align}
\end{subequations}
where the second approximations assume a large EoS $\epsilon$. Note that, sub-horizon, the physical complexity first increases a little simply due to the overall re-scaling caused by the background, but as the numerical example in Fig.~\ref{fig:twofieldcomplexity} shows, this increase is eventually overwhelmed by the growth of the imaginary part of $A_{\delta s \delta s}$ on large scales [cf.~Eq.~\eqref{approxentropycorr}]. We may thus approximate the total growth of complexity during the ekpyrotic phase by $2\sqrt{2}\mathcal{N}_\star$ in the large $\epsilon$ limit.
We immediately notice that this is twice as large as the super-horizon growth of complexity in inflation [recall Eq.~\eqref{eq:CRinfasymsupHN}].
Of course, the total growth depends on the time spent by a mode on sub- and super-horizon scales.
In particular, an entropy fluctuation that exits the horizon very late during ekpyrosis (so that its super-horizon evolution is very short) acquires very little complexity, especially compared to a similar inflationary curvature fluctuation.
This is perhaps the most important difference between ekpyrosis and inflation: the sub-horizon growth of physical complexity is mitigated in ekpyrosis in the large $\epsilon$ limit, while it is significant for inflation; recall Eq.~\eqref{eq:CRinfasymsubHN}.
Another difference is that the physical complexity is further restrained due to the kinetic phase in the ekpyrotic scenario.
Indeed, instead of continuing its growth on super-horizon scales, the complexity saturates to a constant value.\footnote{From the approximate correlator on large scales in the kinetic phase, Eq.~\eqref{eq:Adeltaskinapprox}, one can see that the complexity is dominated by a constant term and a term growing as $\ln([\ln(-k\tau)]^2)$. However, it turns out that as long as $|\Im A_{\delta s\delta s}(\tau_\textrm{e-k})| \gg A_{\delta s\delta s}(\tau_\mathrm{i})$, the constant term dominates over the slowly (logarithmically) growing term. We will see below that a special case where the growing term dominates occurs when $\epsilon$ is very close to $3$ in the ekpyrotic phase.}
It is important to point out, though, that in realistic models the kinetic phase should last only a few $e$-folds before a bounce and reheating occur.

The authors of \cite{Bhattacharyya:2020kgu} conjectured the existence of an upper bound on the growth of complexity in contracting universes. This bound is clearly violated by the entropy perturbation studied here, because their growth depends just as much on the transverse potential as on the background evolution.
Moreover, we would like to point out that analysing the growth of complexity as a function of physical time may be misleading.
We saw that for de Sitter $\dd\mathcal{C}/\dd t\simeq\sqrt{2}H$, but a similar derivative for super-horizon entropy modes in ekpyrosis yields $\dd\mathcal{C}/\dd t\simeq 2\sqrt{2}\epsilon|H|$, assuming $\epsilon$ is so large that the scale factor is approximately a constant and hence $t\approx\tau$.
Thus, one could claim that the growth rate of the complexity of entropy modes in ekpyrosis can be made as large as wanted (it is unbounded by $\epsilon$), and furthermore it is not constant [it keeps growing as the universe contracts since $|H(t)|$ grows].
However, as we saw, the growth rate really is bounded for a fixed number of $e$-folds of ekpyrosis; hence we believe the more appropriate time variable is $\mathcal{N}$.
In that sense, if we think of the growth rate $\dd\mathcal{C}/\dd\mathcal{N}$ as characterising the chaotic nature of the perturbations, then we obtain a hierarchy for the models on large scales as $\dd\mathcal{C}_\mathcal{R}^\mathrm{(inf)}/\dd\mathcal{N}\simeq\sqrt{2}<\dd\mathcal{C}_\mathcal{R}^\mathrm{(ek)}/\dd\mathcal{N}\simeq 2\sqrt{2}$, assuming the entropy fluctuations $\delta s$ in ekpyrosis are later converted into curvature perturbations $\mathcal{R}$.
We will comment on the interpretation of this hierarchy in the discussion section.

\begin{figure}[t]
\begin{center}
\includegraphics[width=0.5\textwidth]{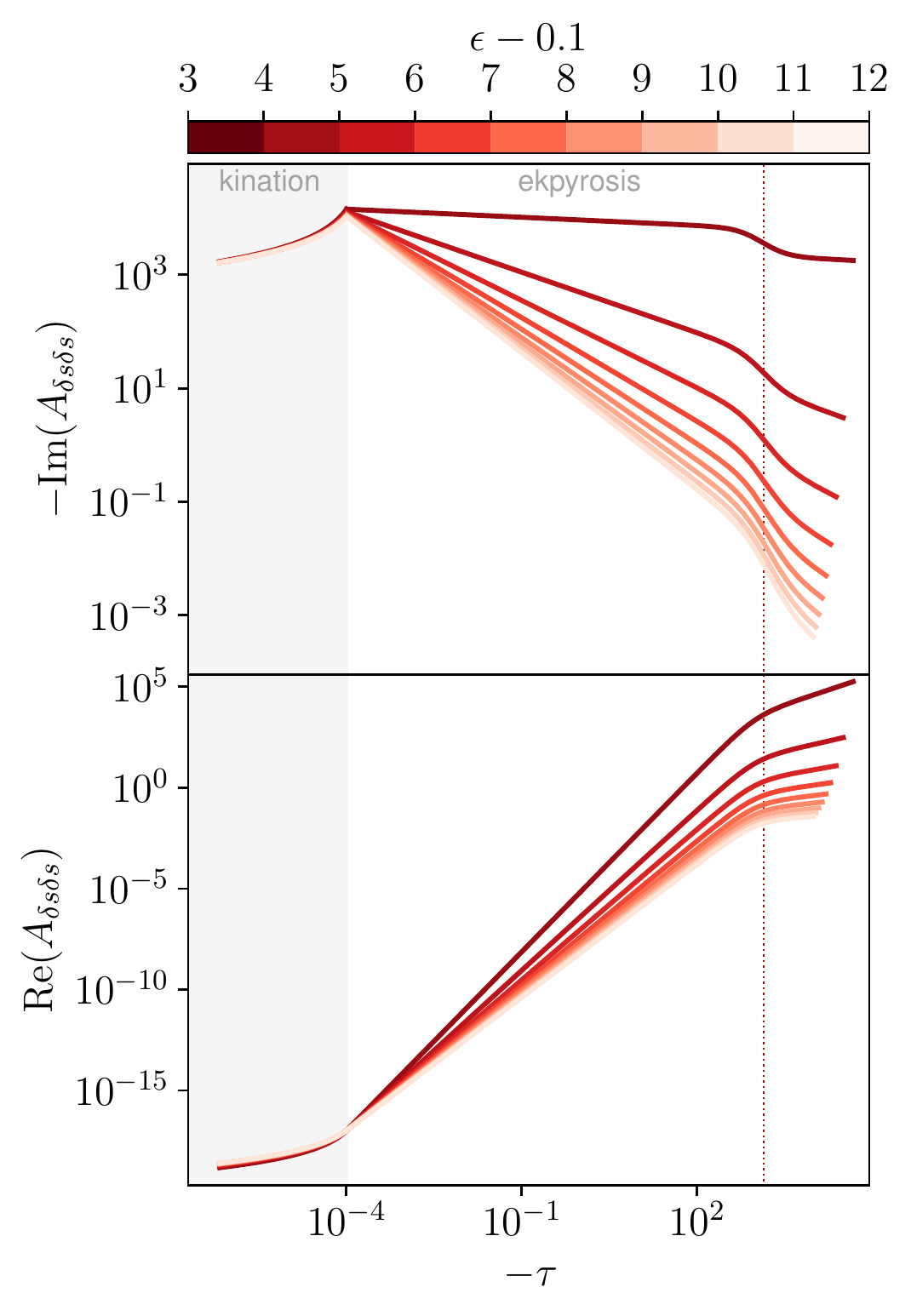}
\caption{Graphs of the real and imaginary parts of the correlator $A_{\delta s \delta s}=a^2A_{ss}$ for the EoS parameter decreasing in unit intervals from $\epsilon = 11.1$ (lightest curve) to $\epsilon = 3.1$ (darkest curve) as a function of conformal time. Here we take $k=0.001$, and $\alpha_s$ is taken to be $3/2$ regardless of the value for $\epsilon$ since the formula for $\alpha_s$ given in the text is not accurate enough for small values of $\epsilon$. The vertical, dotted line is the time of horizon exit. As $\epsilon$ gets smaller, the imaginary part of the correlator grows by less and less (always in absolute value). The numerical evolution has been normalised here such that the correlator takes the same value at ekpyrotic-kinetic matching. In the plot there are always $20$ $e$-folds of ekpyrosis followed by $5$ $e$-folds of kinetic evolution.} \label{fig:smallepscorr}
\end{center}
\end{figure}

\begin{figure}[t]
\begin{center}
\includegraphics[width=0.6\textwidth]{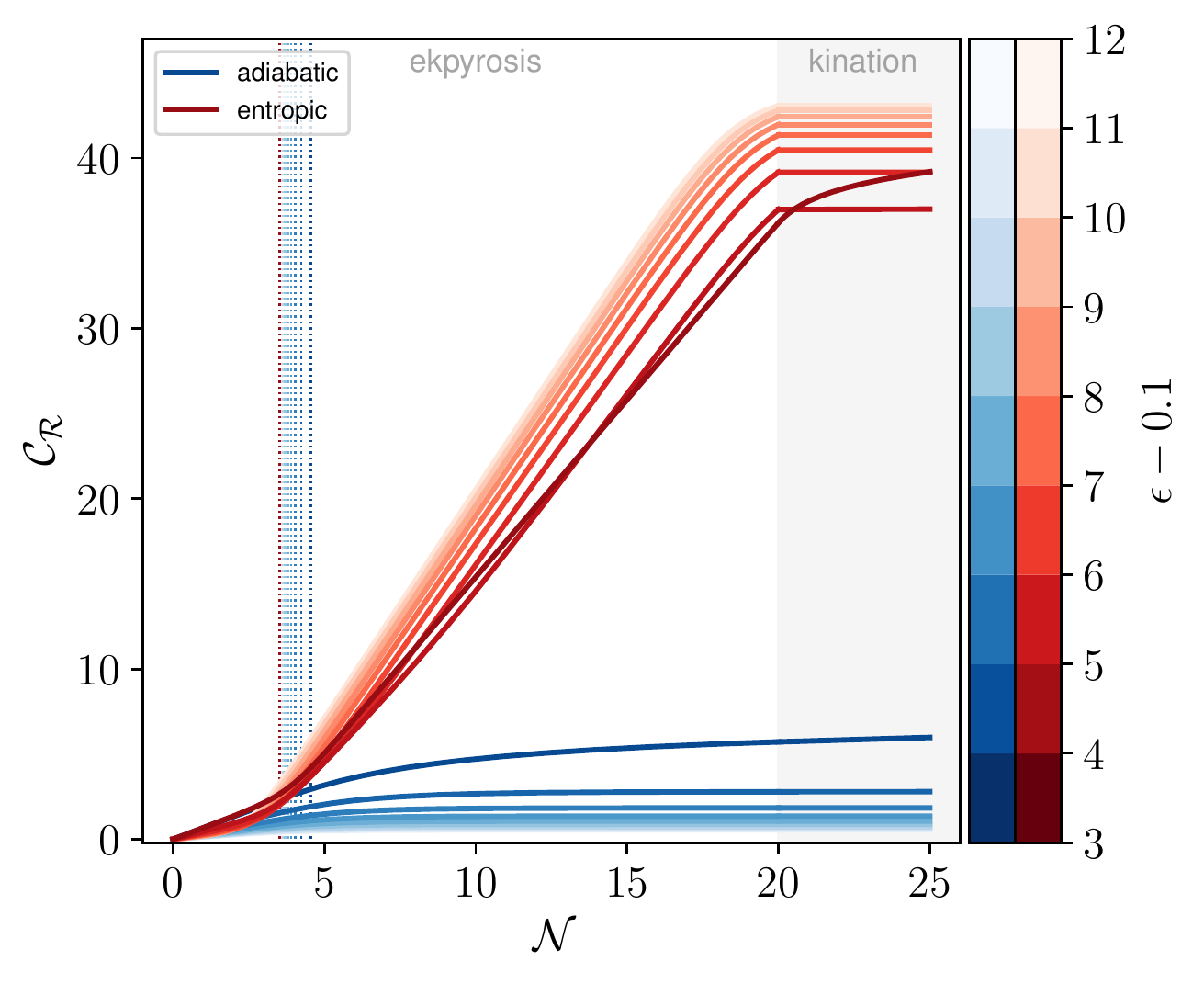}
\caption{The physical complexity for adiabatic (blue) and entropy (red) perturbations as a function of $e$-folding number with the EoS $\epsilon$ ranging from $11.1$ (lightest curves) to $3.1$ (darkest curves) in unit decreases. The same numerical values as in Fig.~\ref{fig:smallepscorr} are used, and the vertical lines are again the times of horizon exit.}
\label{fig:smallepscomp}
\end{center}
\end{figure}

It is interesting at this point to explore a little more the dependence of the complexity in ekpyrosis on the EoS parameter, in particular when it is only marginally in the ekpyrotic domain --- a numerical example of this situation with $\epsilon$ decreasing in unit intervals from  $\epsilon = 11.1$ to $\epsilon = 3.1$ is shown in Figs.~\ref{fig:smallepscorr} and \ref{fig:smallepscomp}. In all cases, the real part of the correlator becomes small, and hence the entropy modes are amplified. However, for small $\epsilon$, the imaginary part grows significantly less. This results in a slightly smaller growth of the complexity of entropy modes.
When the EoS is only marginally larger than $3$, e.g., $\epsilon=3.1$ for the darkest curve in Figs.~\ref{fig:smallepscorr} and \ref{fig:smallepscomp}, the evolution and behaviour of complexity starts deviating: the growth rate is still proportional to $\sqrt{2}/\epsilon$ on sub-horizon scales, but this is now $\mathcal{O}(1)$ rather than being suppressed by a large $\epsilon$; and on super-horizon scales, while the growth rate is now smaller the smaller $\epsilon$ is, the behaviour changes upon the transition to the kinetic phase.
While the complexity saturates (or grows extremely slowly) during kination for $\epsilon$ away from $3$, when $\epsilon$ gets very close to $3$ the complexity continues to grow as $\ln([\ln(-k\tau)]^2)$.
Indeed, as we can see from Fig.~\ref{fig:smallepscorr}, $|\Im A_{\delta s\delta s}(\tau_\textrm{e-k})|$ is actually smaller than $A_{\delta s\delta s}(\tau_\mathrm{i})$ for the darkest curve, so the constant term $(\Im A_{\delta s\delta s})^2/(A_{\delta s\delta s}(\tau_\mathrm{i})\Re A_{\delta s\delta s})$ is suppressed compared to the growing term $A_{\delta s\delta s}(\tau_\mathrm{i})/\Re A_{\delta s\delta s}$ in $X_{\delta s}$ (and correspondingly in $\mathcal{C}_{\delta s}$).
Let us point out that if the imaginary part of the correlator changes too little, it will be difficult to achieve a high degree of classicality of the perturbations. In such extreme cases, one would have to perform a more rigorous calculation of decoherence and the quantum-to-classical transition, along the lines of \cite{Battarra:2013cha}. Models with small background $\epsilon$ are nevertheless easier to construct in supergravity (see, for instance, \cite{Koehn:2013upa} and the discussion in \cite{Lehners:2018vgi}). One should point out, however, that such models require a long ekpyrotic phase, with a very large field displacement, which may be difficult to implement in a reliable effective theory \cite{Lehners:2018vgi} (for a more general discussion, see, e.g., \cite{Ooguri:2006in,Agrawal:2018own,Blumenhagen:2018hsh} and references therein).

The evolution of complexity will change as the universe enters the bounce phase. Just before, during, or just after the bounce, two-field ekpyrotic models envisage a process that uses the entropy perturbations as a source for the adiabatic perturbations. A simple incarnation of this idea, motivated by the original colliding brane ekpyrotic scenario \cite{Lehners:2006ir}, 
is that in the effective theory a bending of the trajectory on scalar field space will occur. Another possibility is that the timing of the bounce itself is modulated by the entropy perturbation, and thus the timing of reheating is modulated \cite{Battefeld:2007st}. 
Whatever the details of the process may be, during this conversion process the adiabatic perturbations (which due to their blue spectrum were essentially absent on large scales) inherit the large-scale properties of the entropic perturbations, in particular the nearly scale-invariant spectrum, the large amplitude, and also the complexity. In this way, large-scale density perturbations are generated at the start of the hot big bang phase. If the universe reaches thermal equilibrium, then the entropy perturbations will be unobservable later on. This in itself constitutes an efficient process of decoherence, rendering the curvature perturbations classical \cite{Battarra:2013cha}. Thus via the so-called entropic mechanism, two-field ekpyrotic models have the potential to explain the observed primordial perturbations --- their complexity being at the root of the later complexities seen in the globally expanding and locally gravitationally collapsing universe.

\section{Matter domination and adiabatic ekpyrosis as other alternatives} \label{sec:matter}

In our review of adiabatic perturbations in section \ref{sec:quantisation}, which was then applied to inflation ($\epsilon\ll 1$) and ekpyrosis ($\epsilon>3$), we always had the underlying assumption that the EoS $\epsilon$ is actually constant (or close enough to a constant as a leading-order approximation).
In more generality, this might not hold, and one may as well construct different early universe scenarios in which the EoS has an important time dependence. The more general equation of motion for the canonical variable of a single adiabatic mode, $v\equiv z\mathcal{R}$, is still given by Eq.~\eqref{FourierEom}, except with $z\equiv a\sqrt{2\epsilon/c_\mathrm{s}}$, $\dd y\equiv c_\mathrm{s}\dd\tau$, and where a prime in the equation of motion now denotes a derivative with respect to the rescaled conformal time $y$.
This way, one allows for a sound speed $c_\mathrm{s}$ possibly different from unity and time dependent, as may arise with a scalar field having a non-canonical kinetic structure.
If $z\propto |y|^n$, then $z''/z=n(n-1)\tau^2$, and the requirement for scale invariance, $z''/z=2/\tau^2$, is achievable if $n=-1$ or $n=2$.
There are thus many possible ways in which $a(\tau)$, $\epsilon(\tau)$, and $c_\mathrm{s}(\tau)$ may yield a scale-invariant power spectrum.
If we consider $c_\mathrm{s}$ to be constant for simplicity, this means $y=\tau$, and so one needs $a\sqrt{\epsilon}\propto\tau^2$ or $a\sqrt{\epsilon}\propto|\tau|^{-1}$.
If $\epsilon$ is also constant, then $a\propto\tau^2$ corresponds to the matter-dominated contracting scenario \cite{Wands:1998yp,Finelli:2001sr,Brandenberger:2012zb} (see also \cite{Quintin:2019pbm} for a recent exposition of the so-called matter bounce scenario and its issues), while $a\propto|\tau|^{-1}$ is slow-roll inflation (de Sitter to this level of approximation); those correspond to the two `standard' adiabatic structure formation scenarios.
Conversely, we could explore cases where $a$ is essentially constant (as in the original ekpyrotic scenario), but where $\epsilon$ is (strongly) time dependent.
In this case, we see that $\sqrt{\epsilon}\propto|\tau|^{-1}$ constitutes an interesting scenario in which the EoS starts small (inflation like, though the universe is very slowly contracting) at early times ($\tau\to-\infty$) and increases to large ekpyrotic-like values at late times ($\tau\to 0^-$).
This is the basis of the adiabatic ekpyrosis\footnote{In a similar spirit, there exist other alternative proposals with time-varying EoS or sound speed (see, e.g., \cite{Khoury:2008wj,Khoury:2010gw,Baumann:2011dt,Joyce:2011ta,Geshnizjani:2011dk,Geshnizjani:2011rm}). One could certainly study the evolution of complexity in any such scenario, but in this section we focus on matter domination and the only other ekpyrotic-like model for concreteness.} model \cite{Khoury:2009my,Khoury:2011ii} (note that in this model adiabatic perturbations behave differently from the adiabatic perturbations in ordinary ekpyrosis studied in sections \ref{sec:ekpyrosis} and \ref{sec:twofields}).

\begin{figure}[t]
\begin{center}
\includegraphics[width=0.5\textwidth]{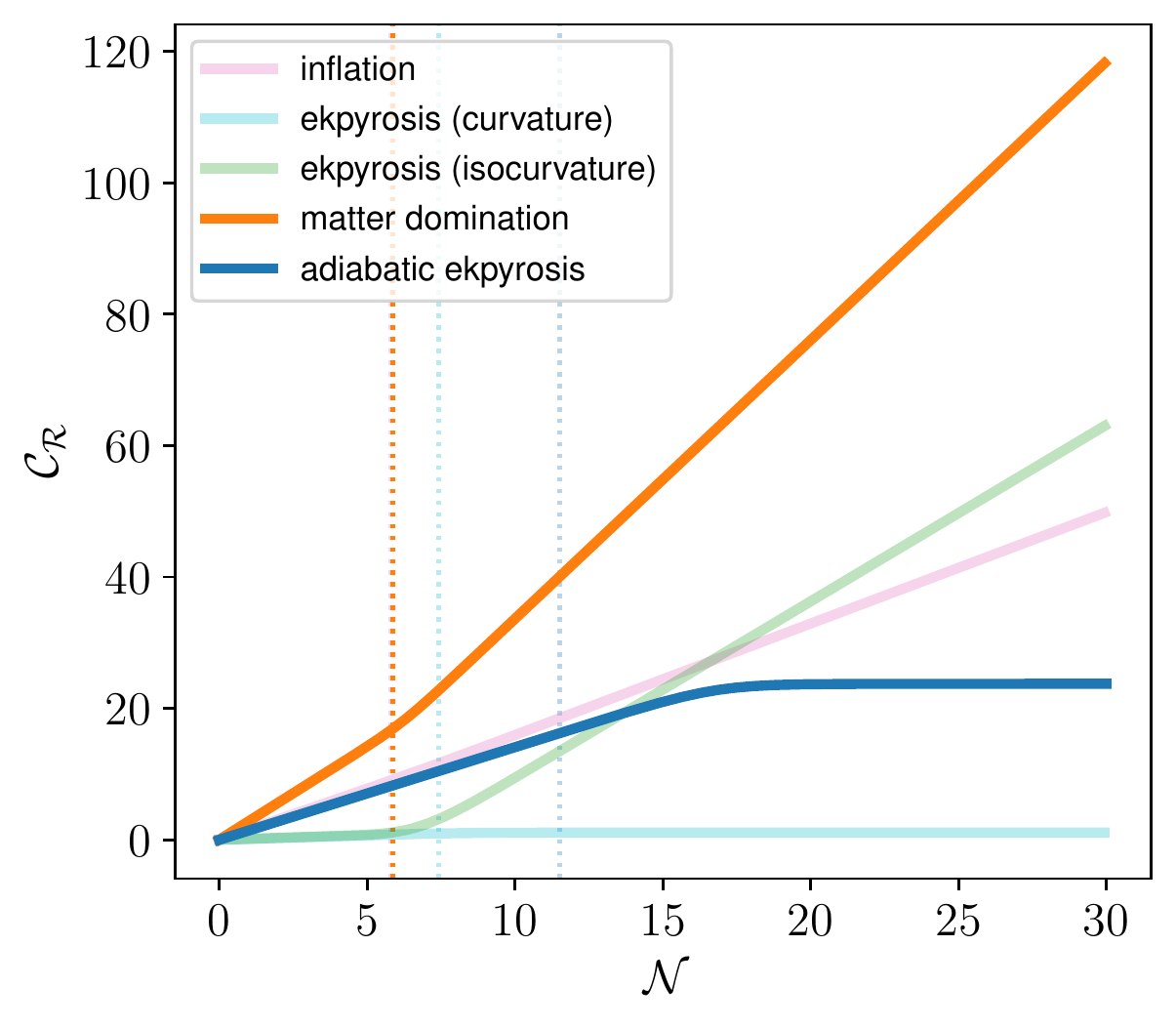}
\caption{Physical complexity (denoted $\mathcal{C}_\mathcal{R}$ in every case) as a function of the number of $e$-folds for matter-dominated contraction (orange curve) and an example of adiabatic ekpyrosis (blue curve), namely using Eq.~\eqref{eq:zadek} for the evolution of $z$. For the latter, the following numerical values are taken: $H_0=-5\times 10^{-4}$, $c=140$, and $k=0.05$. As usual, the vertical, dotted lines indicate the respective horizon exit times. In the adiabatic ekpyrotic model, horizon exit is taken to occur during the phase of rapidly evolving EoS, $z\sim\sqrt{\epsilon}\sim 1/(-t)$. The transition to the phase of constant EoS occurs at later times, when $\mathcal{N}\gtrsim 17$, and from there complexity saturates. The complexity during inflation and two-field ekpyrosis (both curvature and isocurvature modes) with the parameter values of the previous sections are depicted with fainter colours for the sake of comparison.} \label{fig:Call}
\end{center}
\end{figure}

Matter-dominated contraction is an adiabatic scenario with constant EoS $\epsilon=3/2$ [so $a(\tau)\propto\tau^2$]. Thus most expressions of section \ref{sec:quantisation} are immediately applicable, in particular Eq.~\eqref{eq:corralphalarge} for the correlator $A_{\sigma\sigma}$ on large scales with $\alpha_\sigma=3/2$. Multiplying by $z^2=3\bar a_0^2\tau^4$ yields $\Re A_{\mathcal{R}\mathcal{R}}\sim\tau^6$ and $\Im A_{\mathcal{R}\mathcal{R}}\sim|\tau|^5$; hence the amplification is very efficient as $1/\Re A_{\mathcal{R}\mathcal{R}}\sim\tau^{-6}$ and so is squeezing with $\Im A_{\mathcal{R}\mathcal{R}}/\Re A_{\mathcal{R}\mathcal{R}}\sim|\tau|^{-1}$. Regarding complexity, from Eq.~\eqref{eq:CRsubHgen} one has $\mathcal{C}_\mathcal{R}\simeq 2\sqrt{2}\mathcal{N}$ on sub-horizon scales, while the evolution on super-horizon scales can be approximated as $\mathcal{C}_\mathcal{R}(\tau)\simeq\mathcal{C}_\mathcal{R}(\tau_\star)+3\sqrt{2}\ln(\tau_\star/\tau)$; hence the super-horizon complexity growth is $\Delta\mathcal{C}_\mathcal{R}\simeq 3\sqrt{2}\Delta\mathcal{N}$. We note that those are the largest growth rates of complexity encountered so far in this work. We present a full numerical calculation of the complexity during matter contraction in Fig.~\ref{fig:Call} where this is explicit.

Let us now turn our attention to adiabatic ekpyrosis. In a realistic adiabatic ekpyrosis scenario, the phase during which the EoS rapidly evolves can last for only a few $e$-folds at most.
It is then followed by a standard ekpyrotic phase where the EoS settles to a (large) constant value.
This phase can last longer, but perturbations exiting the horizon during this phase will not acquire a scale-invariant power spectrum.
Thus, appropriate model building gives one a scale-invariant power spectrum over the range of modes that are of observational interest, but which becomes blue outside of this range.
Models that can satisfy all the constraints (observational and theoretical \cite{Khoury:2011ii}) can be engineered with some level of tuning.
For the purpose of the present analysis, we will rather consider a toy model from \cite{Khoury:2009my}, which may not meet all constraints, but which will capture the essential features of adiabatic ekpyrosis with regard to complexity.
Specifically, let us parameterise the evolution of the function $z$ as
\begin{equation}
 z=a\sqrt{2\epsilon}=\frac{c(-t)^{2/c^2}}{1+c^2H_0t}\,,\label{eq:zadek}
\end{equation}
where $c$ and $H_0$ are constants.
It is assumed that the scale factor is almost a constant throughout [e.g., $a(t)\propto(-t)^{2/c^2}\approx 1$ with $c\gg 1$], so that physical time and conformal time are approximately equal, $t \approx\tau$.
Then at early times ($\tau\to -\infty$), the EoS rapidly evolves as desired, $\epsilon\sim 1/\tau^2$, while at late times ($\tau\to 0^-$) the EoS tends to a large constant, $\epsilon\simeq c^2/2$.
From this, the function $z$ behaves as in inflation (more precisely the de Sitter limit $\epsilon\to 0$) at early times, $z\sim 1/|\tau|$, which does not only imply the same scale-invariant power spectrum on large scales but also the same complexity growth for that time period on sub- and super-horizon scales, specifically $\mathcal{C}_\mathcal{R}\simeq\sqrt{2}\mathcal{N}$.
Modes of observational interest have thus evolved to large scales when the EoS moves closer to its constant, late-time value.
At that point, we expect the perturbations to be matched with the super-horizon mode solutions of standard single-field ekpyrotic cosmology and acquire the corresponding complexity evolution.
In that limit, $z$ is approximately constant, so from the mode equation in the far infrared ($k\to 0$), $\mathcal{R}_k''+2(z'/z)\mathcal{R}_k'\simeq 0$, we can see that $v_k\sim\mathcal{R}_k\sim c_1+c_2|\tau|$ for some integration constants $c_{1,2}$ that are obtainable upon matching.
Therefore, to leading order the correlators $A_{\sigma\sigma}$ and $A_{\mathcal{R}\mathcal{R}}$ and the corresponding complexities, $\mathcal{C}_v$ and $\mathcal{C}_\mathcal{R}$, all tend to constants at late times.

This can be verified by taking Eq.~\eqref{eq:zadek} and solving the mode equation \eqref{FourierEom} by means of numerical methods.
Subsequently, using the numerical solution for the mode function, one can compute the correlators and then the complexity.
The result is shown in Fig.~\ref{fig:Call} in dark blue.
In the rapidly evolving EoS phase, the complexity behaves as in inflation (light pink curve).
On sub- and super-horizon scales (for $\mathcal{N}\lesssim 11.5$ and $11.5\lesssim\mathcal{N}\lesssim 17$), the curvature complexity $\mathcal{C}_\mathcal{R}$ grows as $\sqrt{2}\mathcal{N}$.
Then, as the EoS tends toward its constant value (i.e., $z\sim\sqrt{\epsilon}\sim\mathrm{const.}$), happening when $\mathcal{N}\gtrsim 17$ (still on super-horizon scales), the complexity tends to a constant; i.e., it saturates.
This is in agreement with the analytical approximation derived in the previous paragraph.

\section{Discussion and conclusions} \label{sec:discussion}

All viable early universe models must in some way be able to explain the fluctuations observed in the CMB. In the present work, we have shown how quantum circuit complexity provides a useful characterisation of the different ways in which cosmological scenarios achieve this goal.
The different theories of the early universe all possess different Hamiltonians (by virtue of having different equations of state), which in turn fully describe the evolution of cosmological perturbations. In that sense, a given model with prescribed initial conditions and Hamiltonian \emph{is} a quantum computer with the necessary complexity to yield the CMB, but we do not know which quantum computer actually evolved our universe. Our calculation precisely extracts the underlying quantum complexity of a given Hamiltonian with its set of initial conditions, given a set of elementary quantum gates. This paper thus addressed the question of how complex a quantum computer \emph{simulating} the evolution of cosmological perturbations of different early universe scenarios would have to be. In essence, we attempted to determine how many quantum gates (taken from a specified set) a table-top experiment would need in order to replicate the transition from initial quantum fluctuations to classical density perturbations.

The models we have analysed (inflation, ekpyrosis, and a contracting matter phase) all rely on the amplification and squeezing of quantum perturbations. But the details of how these phases proceed differ markedly. A useful summary of our results is provided by Fig.~\ref{fig:Call}, and let us also recall the following super-horizon evolutions:
\begin{equation}
 \Delta {\mathcal C}_\mathcal{R}^\textrm{inf.} \simeq \sqrt{2}(1+2\epsilon)\, \Delta {\mathcal N}\,, \quad \Delta {\mathcal C}_{\delta s}^\textrm{ekp.~(entropic)} \simeq 2\sqrt{2}\left(\frac{\epsilon-\frac{3}{2}}{\epsilon-1}\right)\, \Delta\mathcal{N}\,, \quad \Delta {\mathcal C}_\mathcal{R}^\textrm{matter} \simeq 3\sqrt{2}\, \Delta\mathcal{N}\,.
\end{equation}
Two main features are immediately obvious: the complexity that is achieved depends primarily (essentially linearly) on the number of $e$-folds of evolution. And the coefficient of proportionality depends on the cosmological model; i.e., it serves to distinguish the different models.

The growth of complexity is smallest for the most popular early universe model, namely inflation. (In adiabatic ekpyrosis the growth rate is the same, before it caps off.) Thus inflation acts as a comparatively `simple' quantum computer in drawing quantum perturbations out of the vacuum and turning them into effectively classical density perturbations. Contracting models have a higher growth of complexity, and hence they are more `efficient' at quickly producing a complex system; in particular, a contracting matter phase leads to the largest complexity. Ekpyrotic models reside in between inflation and matter contraction, and they possess the distinguishing feature that on sub-horizon scales the growth of complexity is very small, so that essentially the entire complexity comes from super-horizon evolution (cf.~again Fig.~\ref{fig:Call}). It is interesting that the models come out as being so clearly distinguished. Note in particular that within each class of models the specific dependence on the equation of state is rather modest. Also, a dependence on the wavenumber comes about only through its influence on the time of horizon exit, and even this only when there is a significant difference in the growth of complexity before and after horizon crossing. Moreover, there is no explicit dependence at all on the energy scales involved (e.g.~of the potential). Thus complexity provides a truly complementary characterisation of cosmological models, more attuned to their quantum properties.

In order to define complexity we have used a measure that was developed in particular for Gaussian states and that is related to the $\mathrm{Sp}(2,\mathbb{R})$ symmetry of the associated quantum mechanics \cite{Camargo:2018eof,Chapman:2018hou}. This measure is conceptually appealing, as it provides a link with hyperbolic geometry --- see Fig.~\ref{fig:PoincareDisk} for a useful visual illustration of the evolution of cosmological correlations (using the same numerical models as in Fig.~\ref{fig:Call}). Moreover, the hyperbolic measure has a structure that is sensitive to both amplification and squeezing, which are precisely the features that are important for early universe models. A comparison with another popular measure is provided in appendix \ref{app:comparison}. It will be important to see how the present study can be generalised to non-Gaussian corrections, which are bound to play a significant role in future observations.

We have focused on the complexity of the physical perturbations and contrasted it to that of the re-scaled mode functions. This distinction ends up being rather crucial. The canonically normalised mode functions are not directly sensitive to the overall expansion or contraction of spacetime, and thus they miss the sometimes vast changes of physical wavelength that various cosmological models cause (however, they are highly sensitive to the changes in physics occurring near horizon exit, or at junctions with different phases of evolution). Physical perturbations are not only the ones that are directly related to observations, but they depend much more crucially on the entire history of a cosmological phase, and in particular they are sensitive to the conditions at the beginning. Thus the complexity of physical perturbations offers the prospect of better characterising the initial conditions for the cosmological models that are studied, which will be important in terms of incorporating such phases into a complete cosmology. In this respect we suspect that useful links with the puzzles of trans-Planckian perturbations may also be developed in future work.

An additional theoretical avenue that deserves further exploration is the relation of complexity to chaos. For inflation, these issues are already understood to some extent, principally because inflation may be regarded as a thermal system. But for alternative cosmological models, in particular contracting models, such an identification is not available. Yet some chaotic features are certainly present in the dynamics of such models, and it would be interesting to see to what extent complexity may provide a useful diagnostic of these. As is often the case, the confrontation of ideas from different parts of physics is likely to lead to fruitful new insights.

\vskip23pt
\subsection*{Acknowledgments}
We thank Michal Heller and Ro Jefferson for enlightening discussions about quantum circuit complexity.
We gratefully acknowledge the support of the European Research Council (ERC) in the form of the ERC Consolidator Grant CoG 772295 ``Qosmology''.
J.\,Q.~further acknowledges financial support in part from the \textit{Fond de recherche du Qu\'ebec --- Nature et technologies} postdoctoral research scholarship and the Natural Sciences and Engineering Research Council of Canada Postdoctoral Fellowship.

\vskip35pt
\appendix

\section{Comparison with the analytically continued complexity formula}\label{app:comparison}

As mentioned in section \ref{sec:reviewQCC}, a straightforward generalisation of the simplest measure of complexity is obtained by analytically continuing that formula to the case where the frequencies involved may be complex. This approach, and its implications for some cosmological models, has been pursued in \cite{Ali:2018fcz,Bhattacharyya:2020rpy,Bhattacharyya:2020kgu}. In many circumstances, the two approaches yield qualitatively similar results, but there are exceptions.

\begin{figure}[t]
\begin{center}
\includegraphics[width=0.6\textwidth]{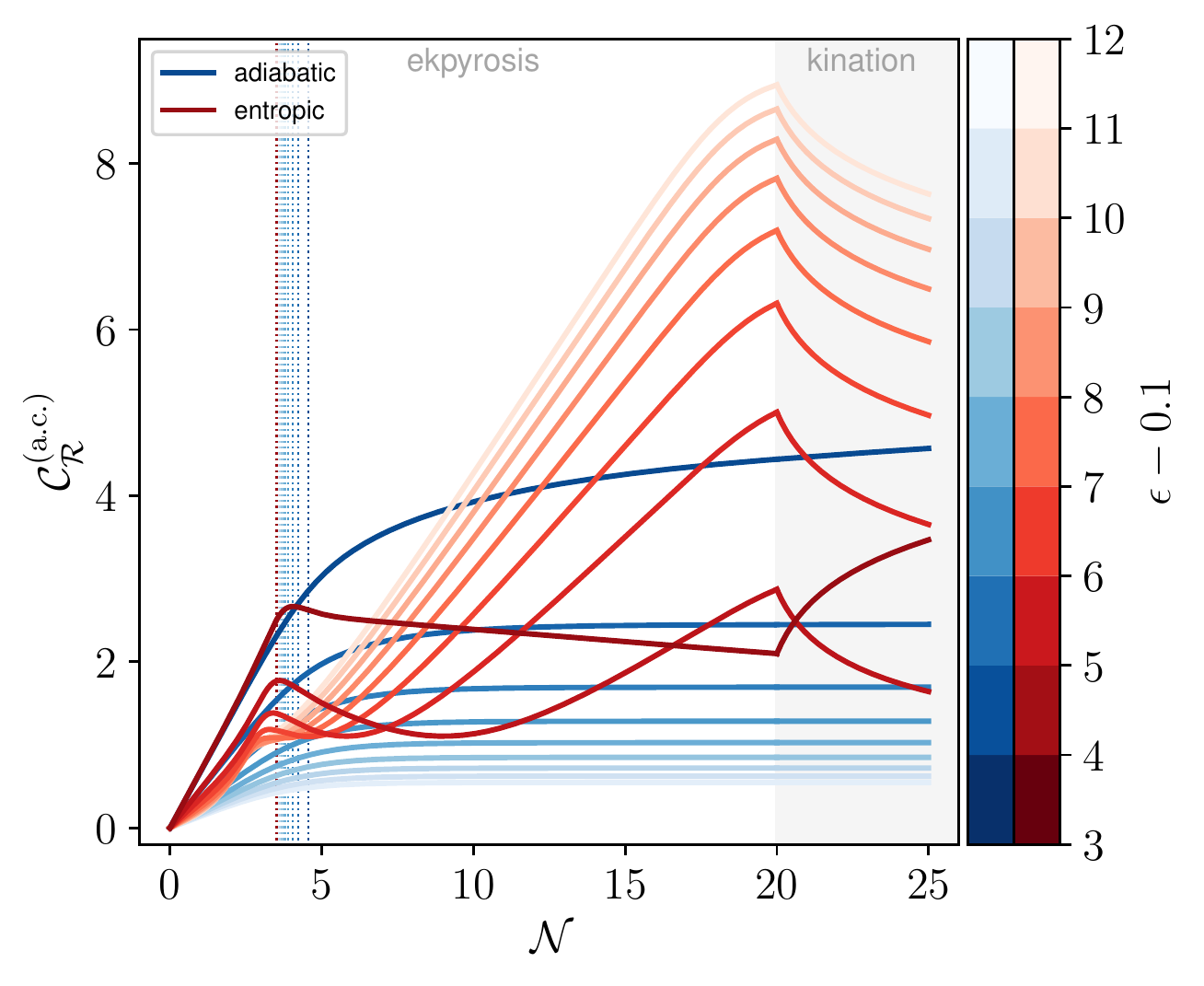}
\caption{The analytically continued complexity for adiabatic (blue) and entropy (red) perturbations as a function of $e$-folding number with the EoS $\epsilon$ ranging from $11.1$ (lightest curve) to $3.1$ (darkest curve) in unit decreases. The models are identical to those in Fig.~\ref{fig:smallepscomp}, where hyperbolic complexity was plotted instead.}
\label{fig:smallepscomp2}
\end{center}
\end{figure}

To illustrate this, it is useful to compare Fig.~\ref{fig:smallepscomp}, showing the evolution of hyperbolic complexity ${\mathcal C}$ for ekpyrotic models with relatively small EoS, with Fig.~\ref{fig:smallepscomp2}, which shows the analytically continued complexity ${\mathcal C}^\mathrm{(a.c.)}$ for the same models. As can be seen from the figures, the evolution of hyperbolic complexity follows a natural progression as the EoS is lowered, while the analytically continued complexity starts showing bizarre features when the EoS approaches the lower bound $\epsilon = 3.$ This may be understood from the following heuristic rewriting of the two definitions of complexity. The inverse of the real part of the correlator determines the amplification $\mathcal{A}$, while the ratio of the imaginary to the real part of the correlator is a measure of the squeezing $\mathcal{S}$. Then the two definitions of complexities may be heuristically written as
\begin{subequations}
\begin{align}
 {\mathcal C} & \sim \frac{1}{\sqrt{2}} \ln \left[ \mathcal{A} + \frac{1}{\mathcal{A}}\left(1+\mathcal{S}^2\right)\right]\,, \\
 {\mathcal C}^\mathrm{(a.c.)} & \sim \frac{1}{\sqrt{2}} \ln \left[\frac{1}{\mathcal{A}^2}\left(1+\mathcal{S}^2\right)\right]\,.
\end{align}
\end{subequations}
Thus we see that the hyperbolic complexity is separately dependent on the amplification and the squeezing and can grow when either of these properties evolves. By contrast, for the analytically continued complexity, amplification and squeezing may counteract each other to some extent. This is exactly what happens for ekpyrotic perturbations when $\epsilon$ is small, since one can see from Fig.~\ref{fig:smallepscorr} that in such a case the imaginary part of the correlator grows only slowly (though it is quite large), while the amplification proceeds without much change compared to the cases with a larger EoS. This has as a consequence that the analytically continued complexity is reduced again, even though the perturbations are both amplified and squeezed. This then leads to the misleading perception that the perturbations evolve just as much during a short phase of kination as during the preceding ekpyrotic phase. The hyperbolic definition avoids this pitfall and seems better suited to us in order to characterise cosmological perturbations.

\newpage
\phantomsection
\addcontentsline{toc}{section}{References}
\let\oldbibliography\thebibliography
\renewcommand{\thebibliography}[1]{
  \oldbibliography{#1}
  \setlength{\itemsep}{0pt}
  \footnotesize 
}
\bibliographystyle{JHEP2}
\bibliography{Complexity}

\end{document}